\documentclass{emulateapj}

\newcommand\gord{K. D. Gordon et al. (2010, in preparation)}

\usepackage{apjfonts}
\usepackage{natbib}
\usepackage{iondefs}

\begin{document}

\title{\textit{Spitzer} Analysis of {\HII} Region Complexes in the Magellanic Clouds:\\Determining a Suitable Monochromatic Obscured Star Formation Indicator}

\author{\sc
B. Lawton\altaffilmark{1},
K. D. Gordon\altaffilmark{1},
B. Babler\altaffilmark{2},
M. Block\altaffilmark{3},
A. D. Bolatto\altaffilmark{4},
S. Bracker\altaffilmark{2},
L. R. Carlson\altaffilmark{5},
C. W. Engelbracht\altaffilmark{3},
J. L. Hora\altaffilmark{6},
R. Indebetouw\altaffilmark{7},
S. C. Madden\altaffilmark{8},
M. Meade\altaffilmark{2},
M. Meixner\altaffilmark{1},
K. Misselt\altaffilmark{3},
M. S. Oey\altaffilmark{9},
J. M. Oliveira\altaffilmark{10},
T. Robitaille\altaffilmark{6},
M. Sewilo\altaffilmark{1},
B. Shiao\altaffilmark{1}
U. P. Vijh\altaffilmark{11}, and
B. Whitney\altaffilmark{12}
}

\altaffiltext{1}{Space Telescope Science Institute, Baltmore, MD 21218, USA; lawton@stsci.edu,
kgordon@stsci.edu}

\altaffiltext{2}{Department of Astronomy, University of Wisconsin-Madison, 475 N. Charter St., Madison, WI 53706, USA}

\altaffiltext{3}{Steward Observatory, University of Arizona, 933 North Cherry Ave., Tucson, AZ 85721, USA}

\altaffiltext{4}{Department of Astronomy, University of Maryland, College Park, MD 20742, USA}

\altaffiltext{5}{Department of Physics and Astronomy, Johns Hopkins University, Baltimore, MD, USA}

\altaffiltext{6}{Harvard-Smithsonian, CfA, 60 Garden St., MS 65, Cambridge, MA 02138-1516, USA}

\altaffiltext{7}{Department of Astronomy, University of Virginia, P.O. Box 3818, Charlottesville, VA 22903-0818, USA}

\altaffiltext{8}{Service d'Astrophysique, CEA/Saclay, l'Orme des Merisiers, 91191 Gif-sur-Yvette, France}

\altaffiltext{9}{Department of Astronomy, University of Michigan, Ann Arbor, MI 48109-1042, USA}

\altaffiltext{10}{School of Physical and Geographical Sciences, Lennard-Jones Laboratories, Keele University, Staffordshire ST5 5BG, UK}

\altaffiltext{11}{Ritter Astrophysical Research Center, University of Toledo, Toledo, OH 43606, USA}

\altaffiltext{12}{Space Science Institute, 4750 Walnut St., Suite 205, Boulder, CO 80301, USA}

\begin{abstract}
\hspace{\parindent} {\HII} regions are the birth places of stars, and
as such they provide the best measure of current star formation rates
(SFRs) in galaxies.  The close proximity of the Magellanic Clouds
allows us to probe the nature of these star forming regions at small
spatial scales.  To study the {\HII} regions, we compute the
bolometric infrared flux, or total infrared (TIR), by integrating the
flux from 8 to 500~$\mu$m.  The TIR provides a measure of the obscured
star formation because the UV photons from hot young stars are
absorbed by dust and re-emitted across the mid-to-far-infrared (IR)
spectrum.  We aim to determine the \textit{monochromatic} IR band that
most accurately traces the TIR and produces an accurate obscured SFR
over large spatial scales.  We present the spatial analysis, via
aperture/annulus photometry, of 16 Large Magellanic Cloud (LMC) and 16
Small Magellanic Cloud (SMC) {\HII} region complexes using the
\textit{Spitzer Space Telescope's} IRAC (3.6, 4.5, 8~$\mu$m) and MIPS
(24, 70, 160~$\mu$m) bands.  Ultraviolet rocket data (1500 and 1900
\AA) and SHASSA {\Ha} data are also included.  All data are convolved
to the MIPS 160~$\mu$m resolution (40 arcsec full width at
half-maximum), and apertures have a minimum radius of $35\arcsec$.
The IRAC, MIPS, UV, and {\Ha} spatial analysis are compared with the
spatial analysis of the TIR.  We find that nearly all of the LMC and
SMC {\HII} region spectral energy distributions (SEDs) peak around
70~$\mu$m at all radii, from $\sim10$ to $\sim400$ pc from the central
ionizing sources.  As a result, we find the following: the sizes of
{\HII} regions as probed by 70~$\mu$m is approximately equal to the
sizes as probed by TIR ($\approx70$ pc in radius); the radial profile
of the 70~$\mu$m flux, normalized by TIR, is constant at all radii
(70~$\mu\mbox{m}\sim0.45~\mbox{TIR}$); the $1\sigma$ standard
deviation of the 70~$\mu$m fluxes, normalized by TIR, is a lower
fraction of the mean ($0.05-0.12$ out to $\sim220$ pc) than the
normalized 8, 24, and 160~$\mu$m normalized fluxes ($0.12-0.52$); and
these results are the same for the LMC and the SMC.  From these
results, we argue that 70~$\mu$m is the most suitable IR band to use
as a monochromatic obscured star formation indicator because it most
accurately reproduces the TIR of {\HII} regions in the LMC and SMC and
over large spatial scales.  We also explore the general trends of the
8, 24, 70, and 160~$\mu$m bands in the LMC and SMC {\HII} region SEDs,
radial surface brightness profiles, sizes, and normalized (by TIR)
radial flux profiles.  We derive an obscured SFR equation that is
modified from the literature to use 70~$\mu$m luminosity,
$\mbox{SFR}~(M_{\odot}~\mbox{yr}^{-1}) =
9.7(0.7)\times10^{-44}~L_{70}~(\mbox{ergs}~\mbox{s}^{-1})$, which is
applicable from 10 to 300 pc distance from the center of an {\HII}
region.  We include an analysis of the spatial variations around
{\HII} regions between the obscured star formation indicators given by
the IR and the unobscured star formation indicators given by UV and
{\Ha}.  We compute obscured and unobscured SFRs using equations from
the literature and examine the spatial variations of the SFRs around
{\HII} regions.
\end{abstract}


\section{INTRODUCTION}\label{intro}
\hspace{\parindent} {\HII} regions are locations of active or recent
star formation where the extreme ultraviolet (UV) radiation from
massive OB stars ionizes the surrounding gas \citep[for a good review
see chapter 5 of][and references therein]{tiel05}.  It is common to
consider {\HII} regions as being large complexes of overlapping
individual {\HII} regions where the sizes and components are not
determined solely by the classical Str\"{o}mgren radius of an
individual star but by an inclusion of all of the physical regimes in
the interstellar medium (ISM) affected by the far-UV photons of the
central sources.  These {\HII} region components include the central
ionizing OB stars, and the surrounding photodissociation regions
(PDRs) and molecular clouds, where the chemistry and the heating are
driven by the UV photons from nearby stars
\citep{tiel05,rela09,wats08}.  Thus, {\HII} regions in this context
are really {\HII} complexes, or more generally, star forming regions.

The observed properties of these star forming regions, including size
and shape, depend on the physical traits such as the number of young
hot ionizing stars, density of neutral gas, abundance of dust, star
formation history, and prior supernovae
\citep[e.g., see][]{hodg74,tiel05,walb02,snid09,harr09}.  These
physical traits are often correlated, and the use of data at many
wavelengths to observe {\HII} regions in different galactic
environments is required to piece them together.

The nature of the hot central OB stars can be studied using UV
\citep[e.g.,][]{smit87,mart05} or nebular emission lines, such as {\Ha}
\citep[e.g.,][]{heni56,hodg83,gaus01}.  The {\Ha} luminosity function
of a galaxy is frequently used to determine many of the physical
properties of {\HII} regions, including the number of ionizing stars,
evolutionary effects, and possible environmental effects
\citep{kenn89,oey98}.  {\HII} region size distributions are directly
related to the luminosity functions and relate to the numbers of
nebula of a given size for a galaxy\citep[e.g.,][]{van81,oey03}.  The
PDRs can be studied by many methods including the [\OI] and [\CII]
infrared (IR) cooling lines and CO radio data \citep[e.g.,][]{kauf99}.
The molecular clouds are typically studied using radio observations of
molecular rotational lines \citep[e.g.,][]{cohe88,genz91,fuku08}.

Present star formation rates (SFRs) are calculated using a tracer of
the UV photons from the young massive stars and spectral synthesis
models \citep[see][]{kenn98}.  {\HII} regions will have some fraction
of their UV photons obscured by dust and some fraction unobscured.
For unobscured SFRs, the ionizing photons can be directly observed via
UV observations or recombination lines such as {\Ha}.  For obscured
{\HII} regions, bolometric IR observations of dust (i.e., the total
infrared (TIR)) can be used to recover the extinguished UV photons.
This is because the dust absorption cross section is highly peaked in
the UV, and the re-emitted flux is in the broad spectral range from
the mid-to-far-IR \citep{kenn98}.  Because the TIR around {\HII}
regions accounts for all of the extincted UV photons, the TIR is
expected to be the single best indicator of SFR obscured by dust.

Combining UV, optical, and IR observations of {\HII} regions across a
whole galaxy allows us to probe the current SFR of that galaxy.  The
Large Magellanic Cloud (LMC) and Small Magellanic Cloud (SMC) are
ideal natural laboratories for studying star forming regions and their
effects on the ISM.  The close proximities of the LMC and SMC, at
$\sim52$ kpc \citep{szew08} and $\sim60$ kpc \citep{hild05},
respectively, allow for detailed star formation studies down to parsec
or sub-parsec scales depending on wavelength.  Any broad study of
{\HII} regions across the LMC and SMC can take advantage of many
multiwavelength observations, including the rocket UV data from
\citet{smit87}, the Southern {\Ha} Sky Survey Atlas (SHASSA) {\Ha}
data from \citet{gaus01}, and the \textit{Spitzer Space Telescope} IR
data from \citet{meix06} and {\gord}.  Furthermore, there are many past
optical {\HII} region surveys that catalog the sources of {\Ha} in
the LMC and SMC \citep[see][]{heni56,davi76,bica95}.

Because of the observational advantages, many researchers rely on a
single-band star formation indicator (i.e., UV, {\Ha}, {\Pa},
8~$\mu$m, 24~$\mu$m, etc).  There are complications in the UV and
optical lines in that they can be greatly impacted by extinction,
thus, requiring extinction corrections.  Another complication is that
the observed UV photons can come from stars of various ages ($< 100$
Myr) and will greatly depend on galaxy type (i.e., quiescent spiral
galaxies, starbursts, etc.; \citep{calz05}).  The 8 and 24~$\mu$m IR
band emission will likely depend on the environment of the host galaxy
because the abundance of the aromatics/small grains that give rise to
their emission depend upon the metallicity and ionizing radiation
present \citep[K. D. Gordon et al. 2010, in
preparation;][]{drai07,drai07b}.  Work done by \citet{calz07} and
\citet{dale05} indicate that 8 $\mu$m makes for a poor star formation
indicator due to large variability of emission in galaxies with
respect to spectral energy distribution (SED) shape and metallicity.
\citet{calz07} also note, along with \citet{calz05}, that a star
formation indicator using 24~$\mu$m by itself can vary from galaxy to
galaxy.  \citet{dale05} claim that SFRs calculated from 24~$\mu$m
emission may be off by a factor of 5 due to variations of 24~$\mu$m
flux with respect to SEDs observed across nearby galaxies.

To compensate for the extinction effects in UV and optical nebular
emission lines, many researchers are now measuring SFRs via a
combination of obscured (TIR, 8~$\mu$m, 24~$\mu$m) and unobscured (UV,
{\Ha}, {\Pa}) star formation indicators
\citep[e.g.,][]{calz07,kenn07,thil07,rela09,kenn09}.  However, the
noted differences in 8 and 24~$\mu$m emission, relative to host galaxy
properties, may still introduce uncertainties to the calculated SFRs
when applying them across large galaxy samples.  In their analysis of
33 galaxies from the \textit{Spitzer} SINGS sample, \citet{calz07}
claim that a combination of {\Ha} and 24~$\mu$m gives the most robust
SFR using a procedure similar to the \citet{gord00} ``flux ratio
method''\footnote{The ``flux ratio method'' uses UV and IR fluxes of
galaxies to derive extinction-corrected UV luminosities
\citep{gord00}}.  Their calibration has a caveat in that it is useful
for actively star forming galaxies where the energy output is
dominated by young stellar populations \citep{calz07}.  \citet{kenn09}
analyze SFRs of nearby galaxies derived by combining {\Ha} with
8~$\mu$m, 24~$\mu$m, and the TIR.  They find that linear combinations
of {\Ha} and TIR provide for the most robust SFRs.

There is little work done on investigating the efficacy of using 70 or
160~$\mu$m as star formation indicators.  \citet{dale05} claim that
the 70~$\mu$m emission may make a good monochromatic obscured star
formation indicator because the 70 to 160~$\mu$m ratio correlates well
with local SFRs.  A \textit{Spitzer} analysis of far-IR compact
sources in the LMC \citep{jacc10} and SMC \citep{jacc10b}, including
compact {\HII} regions, find that the bolometric correction to 70
$\mu$m is modest, due to a typical dust temperature of 40~K.  In a
study of dwarf irregular galaxies, \citet{walt07} find a good
correlation between the brightest 70~$\mu$m regions and optical
tracers of star formation, albeit, with some galaxies contributing
significant 70~$\mu$m diffuse emission at large radii.

Which of the \textit{Spitzer} IR bands most accurately reproduces the
TIR over a large spatial scale?  Employing aperture/annulus
photometry, we analyze 16 LMC and 16 SMC {\HII} region complexes using
the \textit{Spitzer} Infrared Array Camera (IRAC) and Multiband
Imaging Photometer (MIPS) bands.  We determine that the MIPS 70~$\mu$m
band provides for the most accurate monochromatic obscured star
formation indicator based on an analysis of the {\HII} region complex
SEDs, sizes, and radial monochromatic IR fluxes (normalized by the
TIR).  We include an analysis of the spatial distribution of the
unobscured star formation indicators, UV and {\Ha}, relative to the IR
obscured star formation indicators.  We modify an established TIR SFR
recipe from \citet{kenn98} to derive a new monochromatic obscured SFR
equation using the 70~$\mu$m luminosity.

In Section~\ref{sample}, we list the LMC and SMC {\HII} region
complexes sampled in this work.  In Section~\ref{obs}, we explain the
IR, UV, and {\Ha} observations and data reduction.  The analysis of
the multiwavelength photometry is discussed in
Section~\ref{analysis}. The basic results of the photometry, SEDs, and
radial profiles, are discussed in Section~\ref{results} as well as a
discussion of our calculation of the TIR.  Discussions of {\HII}
region normalized radial SEDs, {\HII} complex sizes, normalized radial
profiles, and SFRs are presented in Section~\ref{disc}.  In this
section we also present our derived 70~$\mu$m obscured SFR equation.
We finish with some concluding statements in Section~\ref{conc}.

\section{{\HII} REGION SAMPLE}
\label{sample}
\hspace{\parindent}The 16 LMC and 16 SMC {\HII} regions were selected
by the following three criteria: the center must be peaked in 24
$\mu$m emission, there must be a nearby peak in {\Ha}, and the total
sample of {\HII} regions must sample the full size of the LMC and SMC
as observed in 24~$\mu$m.  Along with sampling {\HII} regions across
the entire LMC and SMC, the 16 LMC and 16 SMC {\HII} regions cover a
wide range of sizes and temperatures.  The final sample size is a
compromise between obtaining good statistics and avoiding overlap
between the large apertures covering each {\HII} region.  Specifics of
the {\HII} regions are listed in Table~\ref{tab:sample}.  The maximum
radii are chosen to include the full extent of the {\HII} region
complex.  We describe how we quantify this in
section~\ref{analysis:phot} and \ref{size}.

\begin{deluxetable}{lcccrc}
\tabletypesize{\scriptsize}
\tablecolumns{6}
\tablewidth{0pt}
\tablecaption{Sample of {\HII} Regions\label{tab:sample}}
\tablehead{
\colhead{}                                     &
\colhead{}                                     &
\colhead{$\alpha$ (2000)}                      &
\colhead{$\delta$ (2000)}                      &
\multicolumn{2}{c}{Maximum Radius}            \\
\colhead{No.}                                   &
\colhead{Name$^a$}                             &
\colhead{(h:m:s)}                          &
\colhead{($^{\circ}$:$\arcmin$:$\arcsec$)} &
\colhead{($\arcsec$)}                          &
\colhead{(pc)$^b$}}
\startdata
\multicolumn{6}{c}{LMC} \\ 
\hline
1      &  N4     & 4:52:08    & -66:55:20    & 1120  & 280  \\
2      &  N11    & 4:56:48    & -66:24:41    & 1505  & 380  \\
3      &  N30    & 5:13:51    & -67:27:22    & 875   & 220  \\
4      &  N44    & 5:22:12    & -67:58:31    & 980   & 250  \\
5      &  N48    & 5:25:50    & -66:15:03    & 1470  & 370  \\
6      &  N55    & 5:32:33    & -66:27:20    & 910   & 230  \\
7      &  N59    & 5:35:23    & -67:34:46    & 875   & 220  \\
8      &  N79    & 4:51:54    & -69:23:29    & 560   & 140  \\
9      &  N105   & 5:09:52    & -68:52:59    & 1785  & 450  \\
10     &  N119   & 5:18:40    & -69:14:27    & 980   & 250  \\
11     &  N144   & 5:26:47    & -68:48:48    & 455   & 115  \\
12     &  N157   & 5:38:36    & -69:05:33    & 1225  & 310  \\
13     &  N160   & 5:39:44    & -69:38:47    & 805   & 205  \\
14     &  N180   & 5:48:38    & -70:02:04    & 1015  & 255  \\
15     &  N191   & 5:04:39    & -70:54:34    & 525   & 130  \\
16     &  N206   & 5:31:22    & -71:04:10    & 1400  & 355  \\
\hline
\multicolumn{6}{c}{SMC} \\ 
\hline
1      &  DEM74  & 0:53:14    & -73:12:18    & 350   & 105  \\
2      &  N13    & 0:45:23    & -73:22:52    & 350   & 105  \\
3      &  N17    & 0:46:42    & -73:31:04    & 385   & 115  \\
4      &  N19    & 0:48:26    & -73:05:59    & 385   & 115  \\
5      &  N22    & 0:48:09    & -73:14:56    & 350   & 105  \\
6      &  N36    & 0:50:31    & -72:52:30    & 525   & 155  \\
7      &  N50    & 0:53:26    & -72:42:56    & 560   & 165  \\
8      &  N51    & 0:52:40    & -73:26:29    & 350   & 105  \\
9      &  N63    & 0:58:17    & -72:38:57    & 350   & 105  \\
10     &  N66    & 0:59:06    & -72:10:44    & 700   & 205  \\
11     &  N71    & 1:00:59    & -71:35:30    & 490   & 145  \\
12     &  N76    & 1:03:43    & -72:03:19    & 350   & 105  \\
13     &  N78    & 1:05:06    & -71:59:36    & 350   & 105  \\
14     &  N80    & 1:08:34    & -71:59:43    & 560   & 165  \\
15     &  N84    & 1:14:05    & -73:17:04    & 980   & 290  \\
16     &  N90    & 1:29:35    & -73:33:44    & 385   & 115  
\enddata
\tablenotetext{\mbox{a}}{Name associated with closest {\Ha} source from \citet{heni56} and \citet{davi76}.}
\tablenotetext{\mbox{b}}{Using the measured distances of $\sim$52,000 pc to the LMC \citep{szew08} and $\sim$60,500 pc to the SMC \citep{hild05}.}
\end{deluxetable}

\begin{figure*}
\includegraphics[angle=0, width=0.98\textwidth]{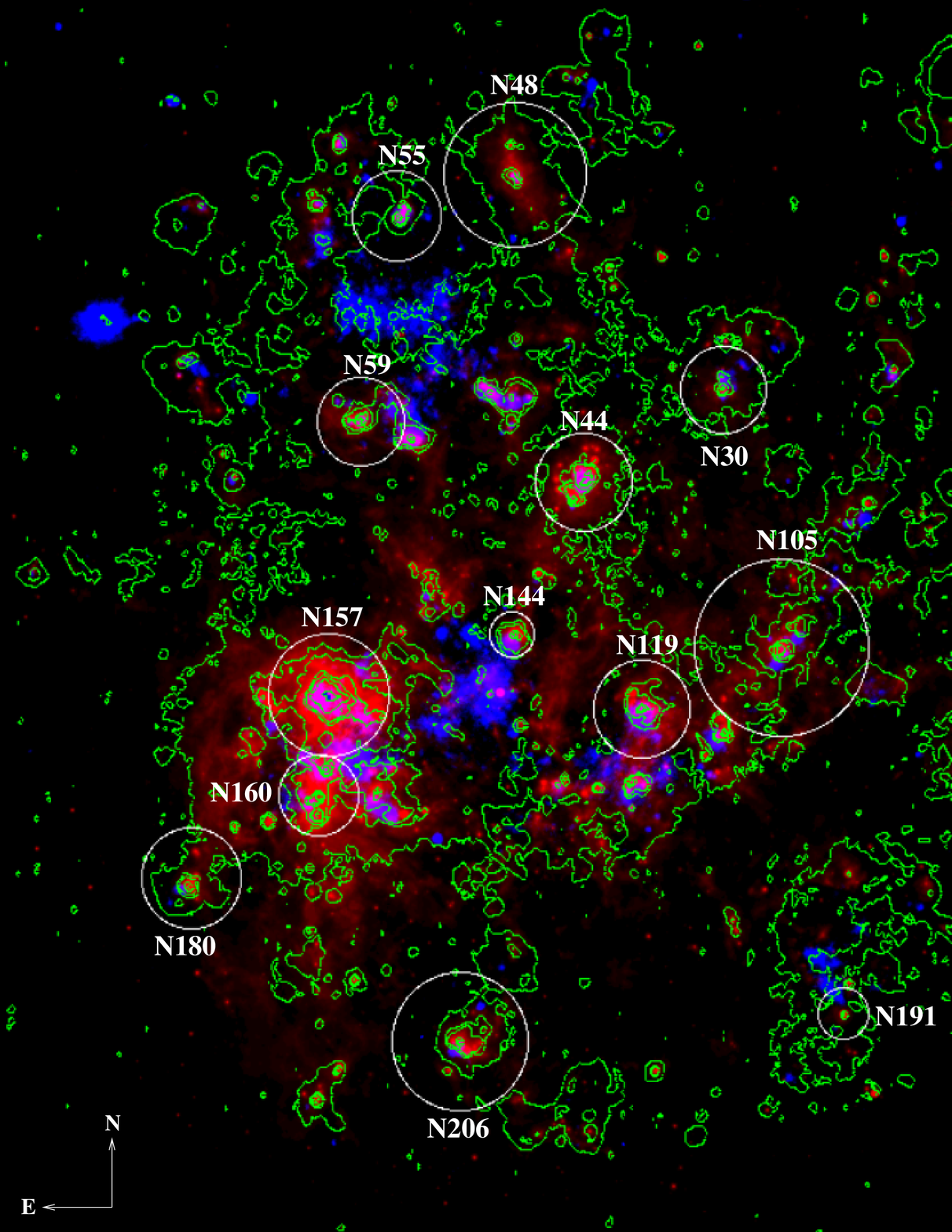}
\caption{LMC image of the UV (blue) from \citet{smit87}, {\Ha} (green
contours) from \citet{gaus01}, and 24~$\mu$m (red) from
\citet{meix06}.  UV and 24~$\mu$m are convolved to the MIPS 160~$\mu$m
resolution ($40\arcsec$).  The SHASSA {\Ha} fluxes are in their
original resolution ($\sim45\arcsec$).  The selected {\HII} regions
are labeled and marked with an aperture that corresponds to the
largest spatial extent for which we measure the photometry (see
Table~\ref{tab:sample}).  The bar in the lower right represents 1 kpc.
North is up, and east is to the left.}\label{fig:LMC_anc}
\end{figure*}

\begin{figure*}
\includegraphics[angle=0, width=0.98\textwidth]{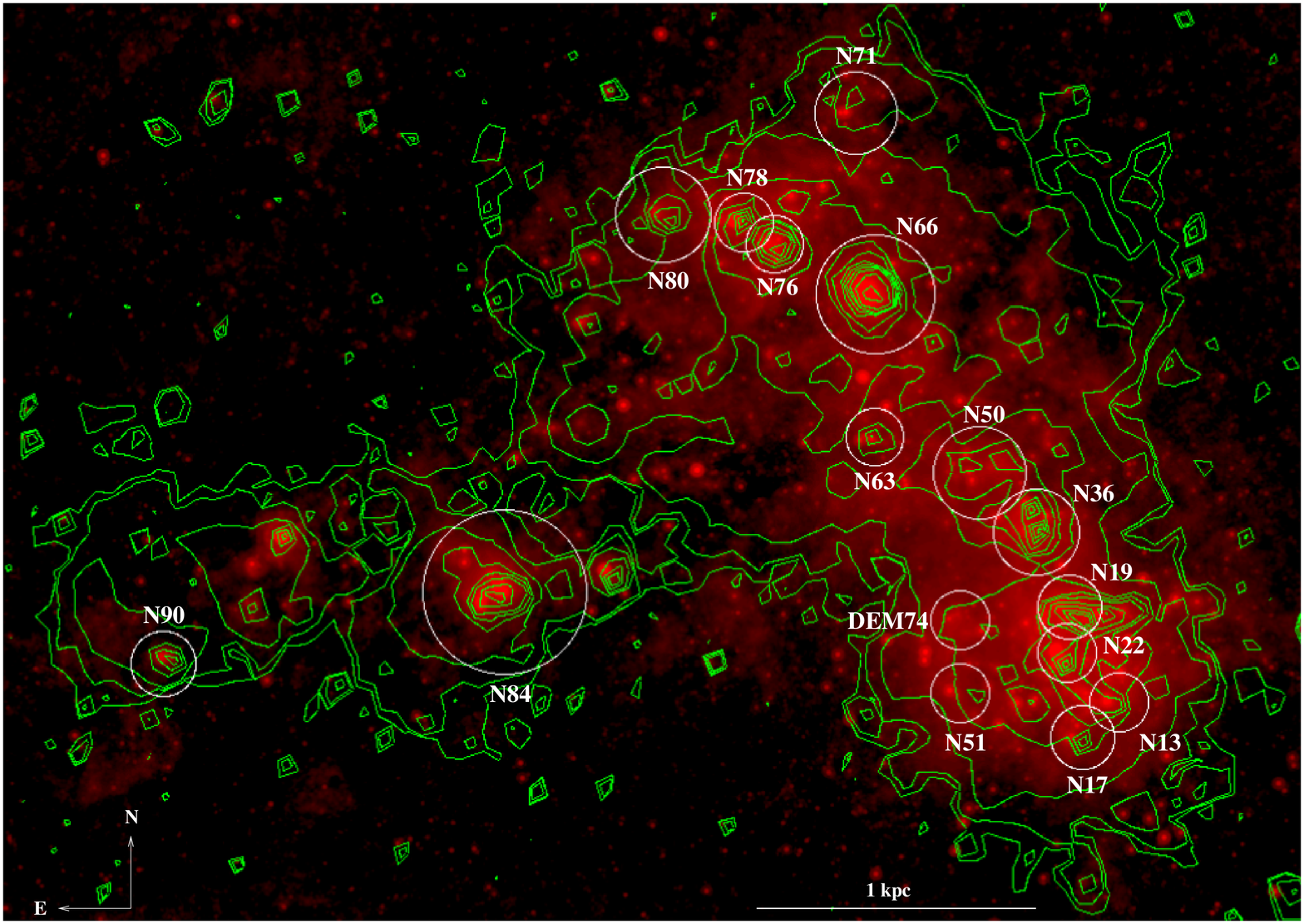}
\caption{SMC image of the {\Ha} (green contours) from \citet{gaus01}
and 24~$\mu$m (red) from {\gord}.  The 24~$\mu$m fluxes are convolved
to the MIPS 160~$\mu$m resolution ($40\arcsec$).  The SHASSA {\Ha}
fluxes are in their original resolution ($\sim45\arcsec$).  The
selected {\HII} regions are labeled and marked with an aperture that
corresponds to the largest spatial extent for which we measure the
photometry (see Table~\ref{tab:sample}).  The bar in the lower right
represents 1 kpc.  North is up, and east is to the
left.}\label{fig:SMC_anc}
\end{figure*}

Shown in Figures~\ref{fig:LMC_anc} and \ref{fig:SMC_anc} are LMC and
SMC images of the unobscured and obscured star formation indicators of
UV (blue--LMC only), {\Ha} (green contours), and 24~$\mu$m (red).  The
{\HII} regions are labeled with circles denoting the maximum extent to
which we measure their photometry (see Columns 5 and 6 in
Table~\ref{tab:sample}).  We do not have UV data for the SMC.  A more
detailed discussion of these figures is presented in
section~\ref{sec:SFR}.  For full \textit{Spitzer} IRAC and MIPS images
of the entire LMC and SMC, see \citet{meix06}, {\gord}, \citet{jacc10},
and \citet{jacc10b}.

\begin{figure*}
\includegraphics[angle=0, scale=0.58]{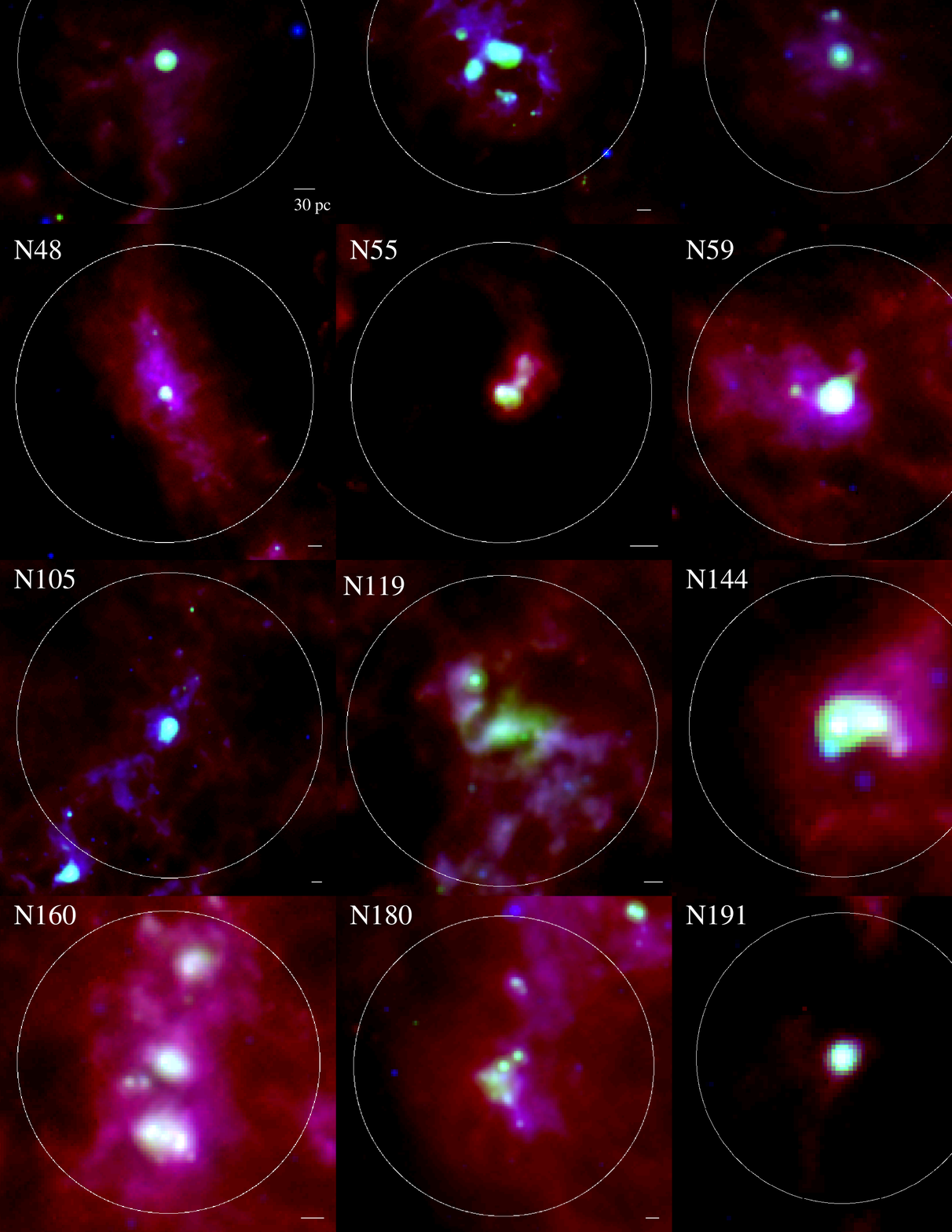}
\caption{Three-color dust images of the 16 LMC {\HII} regions using
the IRAC and MIPS data from \citet{meix06}.  The IRAC 8 $\mu$m is
blue, the MIPS 24~$\mu$m is green, and the coldest dust at the MIPS
160~$\mu$m band is red.  The 8 and 24~$\mu$m fluxes are convolved to
the MIPS 160~$\mu$m resolution ($40\arcsec$).  The bars in the lower
right of each image represent 30 pc.  The white circles represent the
largest apertures used for the photometry (see
Table~\ref{tab:sample}).  North is up and east is to the left.  The
lack of 24~$\mu$m flux in the center of 30~Doradus (N157) is due to
saturation in the core.}\label{fig:LMC_HII}
\end{figure*}

\begin{figure*}
\includegraphics[angle=0, scale=0.58]{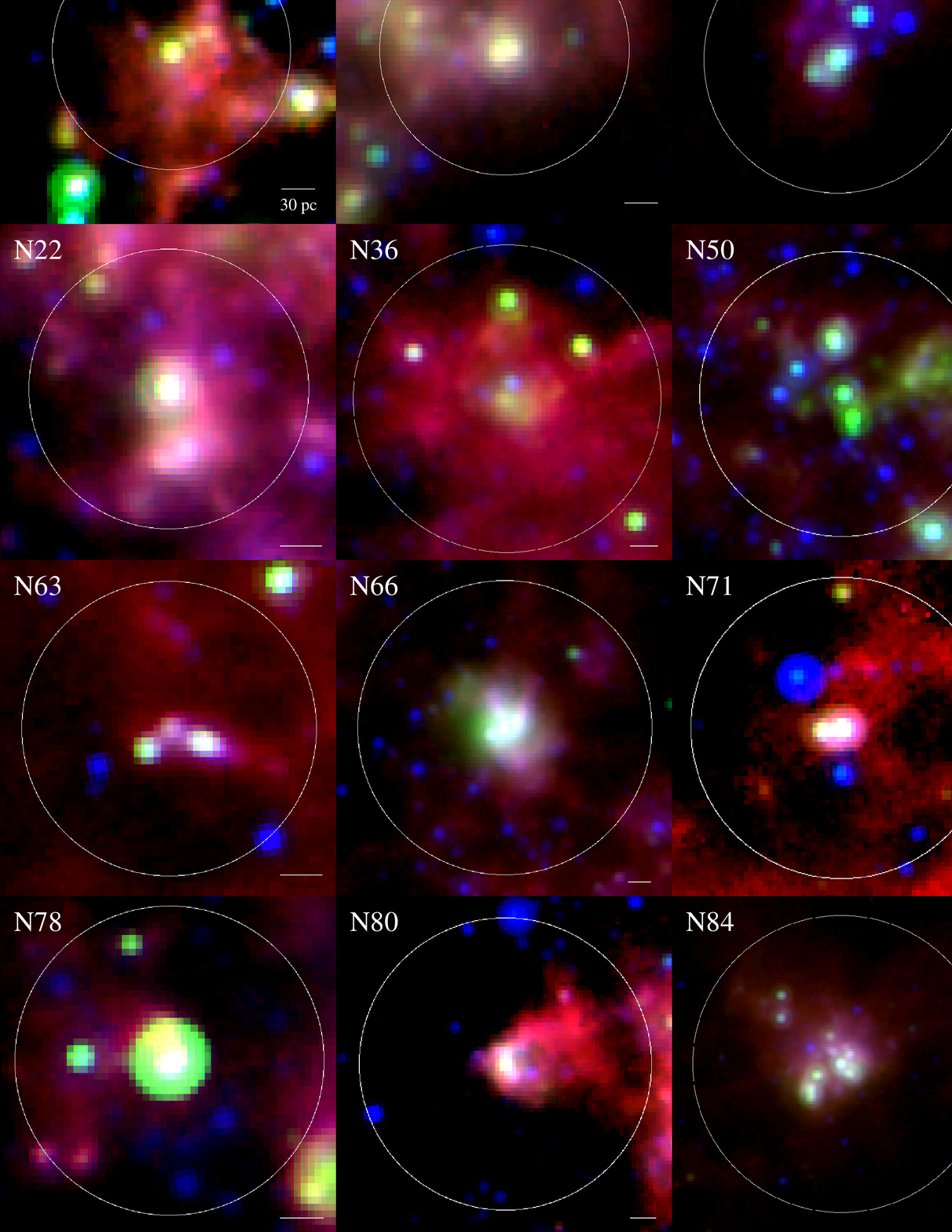}
\caption{Three-color dust images of the 16 SMC {\HII} regions using
the IRAC and MIPS data from {\gord}.  The IRAC 8 $\mu$m is blue, the
MIPS 24~$\mu$m is green, and the coldest dust at the MIPS 160~$\mu$m
is red.  The 8 and 24~$\mu$m fluxes are convolved to the MIPS
160~$\mu$m resolution ($40\arcsec$).  The bars in the lower right of
each image represent 30 pc.  The white circles represent the largest
apertures used for the photometry (see Table~\ref{tab:sample}).  North
is up and east is to the left.}\label{fig:SMC_HII}
\end{figure*}

To view the dust structure of the {\HII} regions, three-color images
of the LMC {\HII} regions are shown in Figure~\ref{fig:LMC_HII}.  The
SMC {\HII} region three-color dust images are shown in
Figure~\ref{fig:SMC_HII}.  In both figures, the aromatic 8~$\mu$m
emission is in blue, the warm 24~$\mu$m dust emission is green, and
the colder 160~$\mu$m dust emission is red.  The sizes of the {\HII}
regions span from 105 pc in radius to 450 pc in radius.  Although we
describe the 8~$\mu$m emission as aromatic, we do not attempt to
remove any possible non-aromatic components.  The IRAC 8~$\mu$m band
may have some contamination from emission of lines such as {\HI},
[\ArII], and [\ArIII] \citep{peet02,lebo07}.  A more detailed
discussion is presented in section~\ref{res:pro}.

\section{OBSERVATIONS}
\label{obs}
\subsection{IR}
\hspace{\parindent}The infrared images are created from the
\textit{Spitzer Space Telescope's} IRAC and MIPS instruments for the
Surveying the Agents of a Galaxy's Evolution (SAGE) project (see
Werner et al. 2004; Meixner et al. 2006; K. D. Gordon et al. 2010, in
preparation).  The observations and data reductions are fully
described in the SAGE LMC overview paper \citep{meix06} and in the
SAGE SMC overview paper (K. D. Gordon et al. 2010, in preparation).

IRAC provides MIR imaging data in four passbands centered around
3.6~$\mu$m, 4.5~$\mu$m, 5.8~$\mu$m, and 8~$\mu$m \citep{fazi04}.  MIPS
provides FIR imaging data in three passbands centered around 24
$\mu$m, 70~$\mu$m, and 160~$\mu$m \citep{riek04}.  The LMC and SMC
IRAC exposures consist of $1\fdg1 \times 1\fdg1$ tiles.  The LMC MIPS
exposures are scan legs $4\,^{\circ}$ long, and the SMC MIPS exposures
are scan legs $2\,^{\circ}-5\,^{\circ}$ long.  The exposure depths are
the same for both galaxies as a result of the same observing
strategies, and variations between the IRAC tiles and MIPS strips have
been removed.  The observations were taken at two epochs separated by
$\sim3$ (LMC/IRAC, LMC/MIPS, SMC/IRAC) and $\sim9$ (SMC/MIPS) months.

The full mosaics, for each IRAC and MIPS band, are created via
processing using the Wisconsin pipeline \citep[IRAC; see description
in][]{meix06} and the MIPS DAT analysis tool \citep[MIPS;][]{gord05}.
The background has been subtracted from the full mosaics of each
galaxy to remove zodiacal and Milky Way cirrus emission.  The mosaics
cover $\sim8\,^{\circ}\times8\,^{\circ}$ for the LMC and $\sim30$
deg$^2$ for the SMC.

The angular resolution of the IRAC mosaics are $1\farcs7$,
$1\farcs7$, $1\farcs9$, and $2\arcsec$ for the 3.6, 4.5, 5.8, and
8~$\mu$m bands.  The angular resolution of the MIPS mosaics are
$6\arcsec$, $18\arcsec$, and $40\arcsec$ for the 24, 70, and
160~$\mu$m bands.  So that the data have equivalent resolution, all of
the IRAC and MIPS point-spread functions (PSFs) have been transformed
to the 40\arcsec~PSF of the MIPS 160~$\mu$m band using the custom
convolution kernels from \citet{gord08}.

\subsection{ANCILLARY DATA}
\hspace{\parindent}The UV data are from imaging taken with
sounding-rocket intrumentation \citep{smit87}.  The bandpasses,
centered around 1495~\AA~and 1934~\AA, are 200~\AA~wide full width at
half-maximum (\fwhm) and 220~\AA~wide {\fwhm}, respectively
\citep{smit87}.  In this paper, we use 1500~\AA~and 1900~\AA~ to refer
to these UV bands.  The 1500~\AA~image has a total exposure time of
114s, and the 1900~\AA~image has a total exposure time of 130s.  The
1500~\AA~PSFs and the 1900~\AA~PSFs have been convolved to match the
160~$\mu$m PSF using the custom convolution kernels of \citet{gord08}.

The {\Ha} images are from the southern sky wide-angle imaging survey
known as the SHASSA \citep[see][]{gaus01}.  The SHASSA survey uses a
Canon camera with a 52 mm focal length lens and an {\Ha} filter
centered at 6563~{\AA} with a 32~{\AA} bandwidth.  The LMC and SMC
images are each comprised of five separate 20m exposures.  The LMC and
SMC are in fields 013 and 010 in \citet{gaus01}, respectively.  SHASSA
images have a similar resolution ($\sim45\arcsec$) to that of the MIPS
160~$\mu$m images ($40\arcsec$).

\section{ANALYSIS}
\label{analysis}
\subsection{Photometry}
\label{analysis:phot}
\hspace{\parindent}The LMC and SMC photometries are measured for the
IRAC (3.6~$\mu$m, 4.5~$\mu$m, 8~$\mu$m) bands, MIPS (24~$\mu$m, 70
$\mu$m, 160~$\mu$m) bands, and the ancillary data for all 32 {\HII}
regions spanning physical sizes from $\sim105$ pc in radius to
$\sim450$ pc in radius (see Table~\ref{tab:sample} and
Figures~\ref{fig:LMC_HII} \&~\ref{fig:SMC_HII}).  The photometry of
the {\HII} regions is measured in concentric annuli from the central
inner $35\arcsec$ aperture outward to the annulus with the largest
radii.  The radii of the annuli are chosen such that they fully sample
the PSF and are large enough to attain good signal-to-noise in the
outer parts of the {\HII} regions where the UV, {\Ha}, and 24~$\mu$m
fluxes quickly drop to background levels.  The flux for a given
annulus is calculated by taking the flux of a larger aperture and
subtracting off the flux from the adjacent smaller aperture.  From
these annuli, we create radial SEDs and radial flux profiles of each
{\HII} region.

The central apertures for each {\HII} region are centered around the
peak 24~$\mu$m flux because the peak 24~$\mu$m flux is observed to
closely coincide with the peak {\Ha} flux \citep{rela09}.  Also, the
24~$\mu$m flux is observed to peak around OB stars at spatial scales
down to less than a parsec \citep{snid09}.  Each 24~$\mu$m peak is
visually checked to make sure that it coincides with a nearby peak in
the {\Ha} SHASSA data and in the {\Ha} catalog of either
\citet{heni56} or \citet{davi76}.  An {\Ha} peak is considered nearby
a 24~$\mu$m peak if it falls within the central three annuli.

The $35\arcsec$ apertures correspond to slightly different physical
scales for the LMC and SMC.  The LMC is $\sim$52 kpc away
\citep{szew08} which corresponds to a physical aperture radius of
$\approx9$ pc.  The SMC is $\sim$60.5 kpc away \citep{hild05} which
corresponds to a physical aperture radius of $\approx10$ pc.  Thus, we
are restricted to measuring the properties associated with the dust
around {\HII} regions at scales larger than $\approx9$ pc for the LMC
and $\approx10$ pc for the SMC.

The largest apertures of the {\HII} region complexes are determined,
by eye, to be where the coldest gas, emitting at 160~$\mu$m, drops in
flux to approximately the level of the background diffuse emission or
where the {\HII} region begins to overlap another {\HII} region.  The
{\HII} region complex sizes are more quantitatively determined in
section~\ref{size}.

To remove the sky background and the local LMC or SMC background, due
to diffuse emission, we perform a background flux removal for each
{\HII} region.  The background flux for every annulus of a given
{\HII} region is computed with an annulus taken at radii of 1.1--1.5
times the radius of the largest aperture.  The flux error for a given
annulus is then the standard deviation of the background sky flux
times the square root of the number of pixels in the annulus.  In the
instances where a neighboring {\HII} region will fall within the
background sky annulus, a mask is created that nulls the pixel values
in the background sky annulus where any large 24~$\mu$m peaks are
observed.  The interloping {\HII} region is nulled in every band.  The
SMC is a smaller galaxy with smaller distances between bright {\HII}
regions, relative to the LMC (see Figures~\ref{fig:LMC_anc} and
\ref{fig:SMC_anc}).  This crowding sets up a lower angular size limit
of the maximum aperture used for many of the SMC {\HII} regions.

We also perform the total IR photometry of each {\HII} region using an
aperture with the largest radius (see Table~\ref{tab:sample}) to
acquire the cumulative flux.  These fluxes are used to plot the {\HII}
region SEDs in section~\ref{res:SEDs}.  For all cumulative fluxes, the
background flux is taken from the same sky annulus as in the
photometry of the individual annuli.

The total IR fluxes of the LMC and SMC galaxies are computed using
aperture photometry with apertures that enclose each galaxy.  The LMC
circular aperture is centered around
5$^{\mbox{h}}$:17$^{\mbox{m}}$:50$^{\mbox{s}}$ right ascension and
-68$^{\circ}$:24$\arcmin$:42$\arcsec$ declination (epoch 2000
coordinates) with a 14,100$\arcsec$ radius.  The SMC aperture is an
ellipse centered around 1$^{\mbox{h}}$:06$^{\mbox{m}}$:00$^{\mbox{s}}$
right ascension and -72$^{\circ}$:50$\arcmin$:46$\arcsec$ declination
(epoch 2000 coordinates) with a major axis radius of 9710$\arcsec$ and
a minor axis radius of 9140$\arcsec$.  The LMC and SMC sky backgrounds
are taken from regions outside of their apertures and away from any
bright IRAC/MIPS infrared sources.

The number of {\HII} regions with a given aperture/annulus decreases
considerably with larger radii for both the LMC and SMC.  Thus, the
conclusions in this work suffer from low number statistics at larger
radii.  For the LMC, the number of {\HII} regions drops to five at an
aperture/outer annulus radius of $1225\arcsec$.  For the SMC, the
number drops to five at an aperture/outer annulus radius of
$525\arcsec$.

\section{RESULTS:}
\label{results}
\subsection{SEDs}
\label{res:SEDs}
\hspace{\parindent}{\HII} region IR SEDs are a combination of the
Rayleigh--Jeans tail of the stellar photospheric emission and emission
from dust grain populations.  The 3.6~$\mu$m fluxes, and to a lesser
degree the 4.5~$\mu$m fluxes, are associated with stellar continua
more than dust \citep{helo04}.  \citet{enge08} find that the
transition from stellar dominated emission to dust dominated emission
in starburst galaxies occurs around 4.5~$\mu$m.  The dust emission at
5.8 and 8~$\mu$m is likely dominated by aromatic emission
\citep{puge89}.  To more conservatively separate the dust emission
from IR emission associated with stellar photospheres, we do not
include the 5.8~$\mu$m band in our analysis
\footnote{The use of the 5.8~$\mu$m band is further complicated by its
relatively low sensitivity among the IRAC bands
\citep{fazi04,meix06}}.

The 24, 70, and 160~$\mu$m FIR emission are associated with silicate
and carbonaceous grains of various sizes, as demonstrated by
interstellar dust grain models \citep[e.g.,][]{drai84,dese90,li01}.
\citet{drai07} find that small grains, of sizes $\sim15-40$~{\AA}, may
contribute a significant portion of the 24~$\mu$m continuum via
single-photon heating.  The 70 $\mu$m emission may be due to a
significant fraction of both small grains and large grains, but
160$\mu$m fluxes are dominated by large grains \citep{dese90}.

From our annulus photometry, we produce IR SEDs for each annulus of
each {\HII} region.  Each of these radial {\HII} region IR SEDs is
created by using the fluxes (mJy~Hz) at 3.6, 4.5, 8, 24, 70, and
160~$\mu$m.  For each {\HII} region, we analyze the SEDs and compare
how they change radially from the core to the outermost measured
annulus.  A full analysis of the radial SEDs is discussed in
section~\ref{disc:seds}.

\begin{figure*}
\includegraphics[angle=0, scale=0.475]{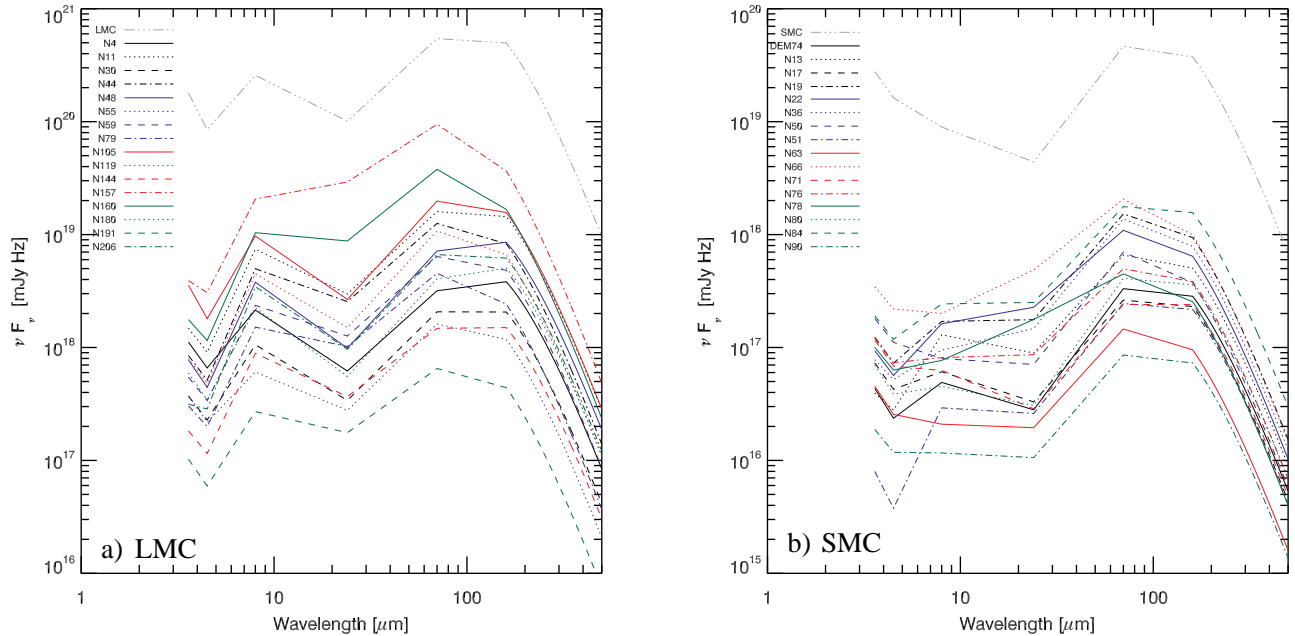}
\caption{IR SEDs of the cumulative fluxes (mJy Hz) of the {\HII}
regions (calculated using the largest aperture) in the --a) LMC and
--b) SMC.  The total galactic LMC and SMC fluxes are also plotted as
triple-dot-dashed lines in gray.  The computation of the
Rayleigh--Jeans tail (160-500~$\mu$m) is explained in
Section~5.2.}\label{fig:SEDs_FULL}
\end{figure*}

We also produce IR SEDs using the cumulative flux of each {\HII}
region and compare these with the SEDs created using the total
galactic LMC and SMC fluxes.  The total LMC and SMC fluxes were
computed using a single aperture around each galaxy (see
section~\ref{analysis:phot}).  The LMC, SMC, and {\HII} region
cumulative fluxes, in Jy, and their $1\sigma$ uncertainties are
tabulated in Table~\ref{tab:HII_fluxes}.  The SEDs are plotted in
Figure~\ref{fig:SEDs_FULL}$a$ for the LMC, and
Figure~\ref{fig:SEDs_FULL}$b$ for the SMC.  The maximum aperture
radius used for each {\HII} region correspond to the apertures marked
in Figures~\ref{fig:LMC_HII} and \ref{fig:SMC_HII} and in the last two
columns of Table~\ref{tab:sample}.

\begin{deluxetable*}{lccccccc}
\tabletypesize{\scriptsize}
\tablecolumns{8}
\tablewidth{0pt}
\tablecaption{IR Fluxes of {\HII} Regions\label{tab:HII_fluxes}}
\tablehead{
\colhead{Name}              &
\colhead{$3.6 \mu$m}        &
\colhead{$4.5 \mu$m}        &
\colhead{$8 \mu$m}          &
\colhead{$24 \mu$m}         &
\colhead{$70 \mu$m}         &
\colhead{$160 \mu$m}        &
\colhead{$T_{\mbox{c}}^{a}$}\\
\colhead{}                      &
\colhead{(Jy)}                  &
\colhead{(Jy)}                  &
\colhead{(Jy)}                  &
\colhead{(Jy)}                  &
\colhead{(Jy)}                  &
\colhead{(Jy)}                  &
\colhead{(K)}}
\startdata
\multicolumn{8}{c}{LMC} \\ 
\hline
LMC$^{b}$   &  $2150\pm40$        &  $1280\pm20$       &  $6860\pm140$     &  $8080\pm320$     &  $127000\pm6400$ &  $265000\pm32000$   & 23.9      \\
N4          &  $13.4\pm0.24$      &  $9.88\pm0.19$     &  $57.3\pm1.2$     &  $49.5\pm2.0$     &  $742\pm37$      &  $2040\pm240$       & 22.6      \\
N11         &  $17.8\pm0.33$      &  $13.6\pm0.3$      &  $198\pm4$        &  $238\pm10$       &  $3730\pm190$    &  $7700\pm920$       & 23.9      \\
N30         &  $4.44\pm0.10$      &  $3.43\pm0.08$     &  $28.2\pm0.6$     &  $27.2\pm1.1$     &  $483\pm24$      &  $1100\pm130$       & 23.4      \\
N44         &  $10.2\pm0.19$      &  $7.71\pm0.15$     &  $133\pm3$        &  $204\pm8$        &  $2930\pm150$    &  $4420\pm530$       & 25.6      \\
N48         &  $9.32\pm0.23$      &  $6.65\pm0.16$     &  $101\pm2$        &  $78.9\pm3.2$     &  $1670\pm80$     &  $4560\pm550$       & 22.6      \\
N55         &  $4.48\pm0.10$      &  $3.27\pm0.07$     &  $16.2\pm0.3$     &  $22.2\pm0.9$     &  $377\pm19$      &  $629\pm75$         & 25.1      \\
N59         &  $6.56\pm0.15$      &  $5.09\pm0.11$     &  $63.0\pm1.3$     &  $101\pm4$        &  $1490\pm70$     &  $2570\pm310$       & 24.9      \\
N79         &  $3.65\pm0.07$      &  $2.99\pm0.06$     &  $40.4\pm0.8$     &  $81.6\pm3.3$     &  $1060\pm50$     &  $1290\pm160$       & 26.9      \\
N105        &  $42.8\pm0.8$       &  $27.0\pm0.5$      &  $259\pm5$        &  $213\pm9$        &  $4620\pm230$    &  $8360\pm1000$      & 24.6      \\
N119        &  $9.59\pm0.23$      &  $7.22\pm0.17$     &  $120\pm3$        &  $120\pm5$        &  $2510\pm130$    &  $3530\pm420$       & 26.0      \\
N144        &  $2.19\pm0.04$      &  $1.73\pm0.03$     &  $23.7\pm0.5$     &  $29.4\pm1.2$     &  $345\pm17$      &  $800\pm96$         & 23.4      \\
N157$^{b}$  &  $46.9\pm0.9$       &  $45.9\pm0.9$      &  $550\pm12$       &  $2340\pm90$      &  $22100\pm1100$  &  $19500\pm2300$     & 29.2      \\
N160        &  $21.2\pm0.4$       &  $17.3\pm0.3$      &  $277\pm6$        &  $702\pm28$       &  $8820\pm440$    &  $8910\pm1070$      & 28.2      \\
N180        &  $7.13\pm0.21$      &  $5.16\pm0.16$     &  $60.0\pm1.3$     &  $44.3\pm1.8$     &  $947\pm47$      &  $2690\pm320$       & 22.4      \\
N191        &  $1.23\pm0.03$      &  $0.892\pm0.021$   &  $7.22\pm0.15$    &  $14.1\pm0.6$     &  $152\pm8$       &  $235\pm28$         & 25.5      \\
N206        &  $3.78\pm0.14$      &  $4.28\pm0.11$     &  $92.2\pm1.9$     &  $77.1\pm3.1$     &  $1550\pm80$     &  $3300\pm400$       & 23.8      \\
\hline
\multicolumn{8}{c}{SMC} \\ 
\hline
SMC         &  $330\pm10$         &  $240\pm10$        &  $240\pm10$       &  $350\pm10$       &  $10800\pm500$   &  $20000\pm2400$     & 24.5      \\
DEM74       &  $0.521\pm0.027$    &  $0.355\pm0.017$   &  $1.31\pm0.03$    &  $2.24\pm0.09$    &  $77.4\pm3.9$    &  $152\pm18$         & 24.2      \\
N13         &  $0.474\pm0.046$    &  $0.420\pm0.034$   &  $3.45\pm0.09$    &  $7.16\pm0.30$    &  $154\pm8$       &  $270\pm33$         & 24.8      \\
N17         &  $0.876\pm0.030$    &  $0.640\pm0.020$   &  $1.63\pm0.04$    &  $2.64\pm0.11$    &  $61.3\pm3.2$    &  $123\pm15$         & 24.1      \\
N19         &  $1.47\pm0.05$      &  $1.09\pm0.03$     &  $4.54\pm0.10$    &  $14.1\pm0.6$     &  $356\pm18$      &  $502\pm60$         & 26.0      \\
N22         &  $1.13\pm0.03$      &  $0.848\pm0.024$   &  $4.33\pm0.10$    &  $18.2\pm0.7$     &  $255\pm13$      &  $343\pm41$         & 26.3      \\
N36         &  $0.937\pm0.043$    &  $0.780\pm0.031$   &  $2.53\pm0.06$    &  $11.8\pm0.5$     &  $321\pm16$      &  $419\pm50$         & 26.5      \\
N50         &  $2.17\pm0.05$      &  $1.67\pm0.04$     &  $2.13\pm0.05$    &  $5.66\pm0.23$    &  $163\pm8$       &  $200\pm24$         & 26.9      \\
N51         &  $0.0960\pm0.0240$  &  $0.0570\pm0.0150$ &  $0.779\pm0.019$  &  $2.08\pm0.08$    &  $57.3\pm2.9$    &  $116\pm14$         & 24.0      \\
N63         &  $0.545\pm0.020$    &  $0.384\pm0.013$   &  $0.558\pm0.013$  &  $1.56\pm0.06$    &  $34.0\pm1.7$    &  $50.9\pm6.1$       & 25.7      \\
N66         &  $4.13\pm0.09$      &  $3.29\pm0.08$     &  $5.34\pm0.11$    &  $38.9\pm1.6$     &  $481\pm24$      &  $531\pm64$         & 27.6      \\
N71         &  $1.39\pm0.03$      &  $1.02\pm0.02$     &  $1.67\pm0.04$    &  $2.29\pm0.09$    &  $56.1\pm2.8$    &  $128\pm15$         & 23.4      \\
N76         &  $1.49\pm0.05$      &  $1.10\pm0.03$     &  $2.17\pm0.05$    &  $6.91\pm0.28$    &  $116\pm6$       &  $204\pm25$         & 24.8      \\
N78         &  $1.20\pm0.03$      &  $0.945\pm0.024$   &  $2.05\pm0.04$    &  $14.2\pm0.6$     &  $105\pm5$       &  $136\pm16$         & 26.5      \\
N80         &  $0.836\pm0.027$    &  $0.579\pm0.019$   &  $1.21\pm0.03$    &  $2.46\pm0.10$    &  $95.6\pm4.8$    &  $191\pm23$         & 24.1      \\
N84         &  $2.29\pm0.06$      &  $1.77\pm0.04$     &  $6.48\pm0.14$    &  $20.1\pm0.8$     &  $413\pm21$      &  $835\pm100$        & 24.0      \\
N90         &  $0.226\pm0.005$    &  $0.177\pm0.004$   &  $0.311\pm0.007$  &  $0.847\pm0.034$  &  $20.0\pm1.0$    &  $38.9\pm4.7$       & 24.2  
\enddata
\tablecomments{Fluxes and derived dust color temperatures for the largest aperture of each
{\HII} region (see Column 5 in Table~\ref{tab:sample}).  $\pm1\sigma$
uncertainties include the statistical and calibration errors summed in
quadrature.  The IRAC 3.6, 4.5, and 8~$\mu$m calibrations have
$1.8\%$, $1.9\%$ and $2.1\%$ uncertainties, respectively
\citep{reac05}.  The MIPS 24~$\mu$m calibration has a $4\%$
uncertainty \citep{enge07}, the MIPS 70~$\mu$m calibration has a $5\%$
uncertainty \citep{gord07}, and the MIPS 160~$\mu$m calibration has a
$12\%$ uncertainty \citep{stan07}.}
\tablenotetext{\mbox{a}}{Dust color temperature calculated using the 70 and
160~$\mu$m fluxes (see section~\ref{SEC:TIR}).}
\tablenotetext{\mbox{b}}{30~Doradus 24~$\mu$m and 160~$\mu$m fluxes are saturated in the core.}
\end{deluxetable*}

The total LMC SED we plot in Figure~\ref{fig:SEDs_FULL}$a$ (gray
triple-dot-dashed line) is very similar, in terms of relative
strengths between bands, to that produced in \citet{bern08}.  The
total SMC SED we plot in Figure~\ref{fig:SEDs_FULL}$b$ (gray
triple-dot-dashed line) is consistent with the SMC SED in {\gord}.
Nearly all of the LMC/SMC {\HII} regions in our sample peak around
70~$\mu$m.  The few {\HII} regions where the 160~$\mu$m flux exceeds
that of the 70~$\mu$m flux, in particular N4, N44, and N180 in the
LMC, have maximum aperture sizes that are among the largest in the
sample.  The paucity of aromatic emission in the SMC, relative to the
LMC, is further discussed in section~\ref{radial}.

Our results match well with those of \citet{degi92}, who observed six
{\HII} regions in the LMC using \textit{IRAS} and noted the peak
fluxes to be near 60~$\mu$m.  Other studies have found peak
wavelengths of {\HII} region SEDs at similar wavelengths, including
the results of \citet{inde08}, who use models of typical {\HII}
regions and find the peak emission between 40 and 160~$\mu$m,
depending on the model used.  Similar work has been done on entire
galaxies.  From modeling and observations, \citet{dale01} and
\citet{dale05} show that nearby galaxies have IR SEDs that peak
between 40 and 160~$\mu$m.

The large fluxes of the 30~Doradus (N157) SED in
Figure~\ref{fig:SEDs_FULL}$a$ (red dot-dash line) are due to an
abundance of hot stars.  30~Doradus is the largest {\HII} region in
the Local Group \citep{walb02} at $\approx200$ pc in size
\citep{rubi98}, and contains more than 100 OB stars \citep{walb97}.
Many of the stars are of type O3, which are among the hottest, most
luminous stars known \citep{mass98}.  30~Doradus is the brightest
{\HII} region in our sample at approximately an order of magnitude
fainter than the total cumulative LMC flux.  There is no comparably
bright {\HII} region in our sample, other than perhaps N160, which is
spatially located near the 30~Doradus complex.  An analysis of
30~Doradus in the FIR by \citet{agui03} finds that this {\HII} region
contains $\approx20\%$ the total LMC FIR emission.  The brightness we
compute is a lower limit because both the 24~$\mu$m and 160~$\mu$m
fluxes are saturated in the core.  The saturation in the central
annuli at 24~$\mu$m artificially lowers the surface brightness we
measure out to $\sim35$ pc.  The saturation in the central aperture of
the 160~$\mu$m emission lowers the surface brightness we measure out
to $\sim9$ pc.  We do not include 30~Doradus in our quantitative
analysis of star formation indicators, but we do include a brief
qualitative discussion of this {\HII} region in section~\ref{radial}
in order to highlight some of the effects a hotter {\HII} region may
have on our conclusions.

The SMC {\HII} region IR SED fluxes are about an order of magnitude,
or more, fainter than the LMC {\HII} region IR SED fluxes.  There are
several possible explanations for this.  For example, metal poor
galaxies will have relatively less dust extinction giving rise to a
greater interstellar radiation field and less IR emission, or an
abundance of very small grains/aromatics can preferentially increase
the Wien's side of the SED \citep{gall05,gala09}.  However, there are
systematically fewer ionizing photons (\Ha) \textit{and} fewer
reprocessed IR photons (TIR) in the SMC {\HII} regions relative to the
LMC {\HII} regions (see Section~\ref{sec_SFRrad} and
Figure~\ref{fig:SFR_Ha}).  TIR and {\Ha} are good star formation
indicators \citep{kenn98}, and decrements in both of them are evidence
that intrinsically lower SFRs are occurring in the SMC {\HII} regions.

This can also be understood by considering the luminosity functions
for {\HII} regions.  The {\Ha} luminosity function follows a power law
where the number of {\HII} regions with a given range of luminosities
is proportional to $L^{-\alpha}$ \citep[e.g.,
see][]{kenn80,kenn86,oey98}.  In general for galaxies, the shape of
the {\Ha} luminosity function depends on the number of ionizing
photons and the effects from evolution \citep{oey98}.  \citet{kenn86}
find an $\alpha\sim1.65\pm0.15$ for the LMC and an
$\alpha\sim1.75\pm0.15$ for the SMC using $\sim200$ LMC and $\sim100$
SMC {\HII} regions from the catalog of \citet{davi76}.  Recent work
from \citet{giel06} finds that the LMC and SMC have power laws of
$\alpha\sim2$.  There is no universal upper cutoff for {\HII} region
{\Ha} luminosity functions, so galaxies with higher SFRs will
statistically have more luminous {\HII} regions.  Between the LMC and
the SMC, \citet{kenn89} find more of the most luminous {\HII} regions
in the LMC, and \citet{kenn86} explicitly show the LMC {\Ha}
luminosity function cuts off at larger luminosities, which is not
surprising considering the presence of 30~Doradus.  There are few
calculations of {\HII} region IR luminosity functions, so it is
uncertain how the {\Ha} luminosity function compares with the IR
luminosity function.  However, in one such study, \citet{liva07}
compare the frequency distribution of \textit{IRAS} 100~$\mu$m
emission for LMC and SMC {\HII} regions, and find that the LMC {\HII}
regions have a higher IR luminosity cutoff (see their Figure~4).

\subsection{TIR}\label{SEC:TIR}
\hspace{\parindent}The bolometric IR flux, TIR, is a measure of the
total dust emission in the IR.  TIR makes for a sensitive tracer of
obscured star formation around {\HII} regions because the absorption
cross section of dust is highest in the UV \citep{kenn98}.  We
calculate TIR for each {\HII} region annulus, the cumulative {\HII}
region apertures, and the total LMC and SMC galaxy apertures.  Our
calculation of TIR includes the 8, 24, 70, and 160~$\mu$m fluxes, as
well as the Rayleigh--Jeans tail, from 160 to 500~$\mu$m, of a
computed modified blackbody (MBB) function.  We do not use the 3.6 or
4.5~$\mu$m bands so as to avoid contributions from stellar continua.

The TIR calculation is done by integrating the area under the linear
interpolation of the flux (from 8 to $160~\mu$m), in flux ($F_{\nu}$)
versus frequency ($\nu$) space, then adding the far-IR Rayleigh--Jeans
tail of the MBB.  The total flux under the Rayleigh--Jeans tail is
computed by integrating under the MBB, in flux ($F_{\nu}$) versus
frequency ($\nu$) space, from 160 to $500~\mu$m and using frequency
steps equivalent to 0.25~\AA.

The MBB is defined as
\begin{equation}\label{EQ:MBB}
MBB = \chi~{\nu}^{\beta}~B_{\nu}(T),
\end{equation}
where $\chi$ is a constant, $\beta$ is the emissivity index, and $B$
is the Planck function (mJy) with dust color temperature $T$ (K).  The
shape of the MBB component is calculated using the ratio of the 70 and
160~$\mu$m fluxes to numerically derive a temperature of the
dust\footnote{The larger grains associated with the 70 and 160~$\mu$m
fluxes are least affected by single-photon heating and, thus, produce
a more accurate Planck function.}, and a $\beta = 2$.  We choose the
classical emissivity index of $\beta = 2$ \citep{geza73,drai84} for
simplicity.  The emissivity index depends upon the chemical nature and
internal physics of the grains and is generally thought to be from $1
\leq \beta \leq 2$ \citep{dupa03} with some indications that it can be
greater than two \citep{menn98,dese08}.  Determining beta is
problematic because $T$ and beta are degenerate.  Also, flux errors
can make determining beta more problematic \citep{shet09}.

The uncertainty in TIR is calculated via standard error propagation
and using the uncertainties in the 8, 24, 70, and 160~$\mu$m bands as
well as the uncertainty in the MBB.  The uncertainty in the MBB
component is calculated via a Monte Carlo method.  Random 70 and
160~$\mu$m fluxes are generated assuming Gaussian distributions about
the measured fluxes with an $\hbox{{\fwhm}}\approx2.35\sigma_F$.  The
new MBB dust color temperature is then calculated using the ratio of
the new 70 and 160~$\mu$m Planck functions.  The flux from 160 to
500~$\mu$m is integrated for this new MBB component as described
before.  This process is repeated 1000 times, and the standard
deviation of the integrated fluxes is the $1\sigma$ error in our MBB
measurement.  The MBB uncertainty is assumed to be independent of the
other flux errors despite being dependent upon the 70 and 160~$\mu$m
fluxes.  Thus, our final quoted TIR uncertainties are slightly larger
than the true TIR uncertainties.  There are additional uncertainties
not taken into account in our calculations such as differences in the
emissivity index.  We also do not know the contribution of
single-photon heating to the 70 or 160~$\mu$m fluxes that might lead
us to calculate a temperature that is too high.

The MBB component (160--500~$\mu$m) is not the dominant contributor to
the IR bolometric flux because all of the SEDs are peaked blueward of
160~$\mu$m (see sections~\ref{res:SEDs} and \ref{disc:seds}).  The
hottest {\HII} region is 30~Dor (N157) at 29.2 K, which puts the peak
of the MBB at $\sim100$~$\mu$m.  The coolest {\HII} region is N180 at
22.4 K, which corresponds to a peak in the MBB at $\sim130$~$\mu$m.
The LMC {\HII} regions are generally cooler than the SMC {\HII}
regions, using the cumulative fluxes from Table~\ref{tab:HII_fluxes}.
However, the LMC cumulative fluxes are measured using larger apertures
which introduce a greater percentage of flux from cold dust at greater
radii.

The individual annuli of each {\HII} region exhibit a wider range of
temperatures than the cumulative fluxes in Table~\ref{tab:HII_fluxes}.
The LMC {\HII} region annulus with the warmest dust is around 30~Dor,
with a color temperature of 41.5 K and a peak in the MBB at
$\sim70$~$\mu$m.  The LMC {\HII} region annulus with the coldest dust
is around N180, with a color temperature of 21.3 K and a peak in the
MBB at $\sim135$~$\mu$m.  The SMC {\HII} region annulus with the
warmest dust is around N36, with a color temperature of 39.2 K and a
peak in the MBB at $\sim75$~$\mu$m.  The SMC {\HII} region annulus
with the coldest dust is around N90, with a color temperature of 19.1
K and a peak in the MBB at $\sim150$~${\mu}$m.  The annuli with the
warmest dust are always the central aperture, except for N36, N50, and
N63 in the SMC.  These {\HII} regions have the warmest dust in annuli
just outside of the center aperture.  The coldest dust for each {\HII}
region is always found at the outer annuli.  The SEDs of the {\HII}
region annuli are further discussed in Section~\ref{disc:seds}.

For the LMC {\HII} regions, the MBB integrated
component ranges from $\sim2\%$ of the $8-160~\mu$m integrated fluxes
for the inner annuli of 30~Doradus to $\sim25\%$ for the outer annuli
of the cooler {\HII} regions (N48, N180, and N206).  For the SMC
{\HII} regions, the percentage of the MBB integrated fluxes to that of
the $8-160~\mu$m integrated fluxes range from $\sim4\%$ (inner
aperture of N66) to as high as $\sim30\%$ for the coldest {\HII}
region outer annulus (N90).  

If the peak of a blackbody or MBB shifts slightly redward or blueward,
the intensity near the peak will not significantly change.  However,
even slight shifts in the peak of the blackbody will greatly affect
the intensities on the far Wien or Rayleigh--Jeans sides of the
blackbody.  As noted above, all of the {\HII} region SEDs in our
sample peak between $\sim70$ and $150~\mu$m, and as discussed in
Section~\ref{radial}, the 70~$\mu$m emission is $\approx40\%-50\%$ of
the TIR, on average, for both the LMC and SMC at all radii measured.
Therefore, the MBB Rayleigh--Jeans component we measure is greatly
dependent upon the 70~$\mu$m flux.  The spreads in the peak of the
SEDs give rise to a spread in the relative importance of the
Rayleigh--Jeans tail to the TIR.  The fluxes in the Rayleigh--Jeans
tails of the MBBs, relative to the TIR, range from $\sim1.7\%$, for a
dust color temperature of 41.5~K, in the central annuli of 30~Doradus,
to $\sim25\%$, for a dust color temperature of 19.1~K in the outer
annuli of N90.

Our defined TIR ($8-500~\mu$m) is similar to other calculations of IR
bolometric flux in the literature, except we do not go quite as far
into the far-IR as some \citep[e.g., 8--1000~$\mu$m used
in][]{kenn98}.  Most of the IR flux around obscured star forming
regions will emit in the $10-120~\mu$m range \citep{kenn98}.  We
believe our TIR results are robust; we do not assume a single SED
parameterization but instead compute the TIR for each individual
{\HII} region annulus.  Our TIR values are nearly identical to what
are predicted by \citet{dale02} for most of the annuli around our
{\HII} regions.  The \citet{dale02} formalism (see Equation~(4) in
their work) is derived using \textit{IRAS} and \textit{Infrared Space
Observatory \mbox{(}ISO\mbox{)}} observations of normal galaxies and
using the 60 to 100~$\mu$m flux ratios to parameterize the range of
possible galaxy SEDs.  Our results differ by a factor of a few in the
outer annuli of our {\HII} regions.  This is not unexpected because
the cold dust at 160~$\mu$m contributes more at these radii.
\citet{dale02} make a point of noting that their TIR results will be
less accurate for colder galaxies because the parameterization is
calculated using the \textit{IRAS} 100~$\mu$m band which does not
extend to long enough wavelengths to include the flux from the coldest
dust.

\subsection{Morphology and Radial Surface Brightness Profiles}
\label{res:pro}
\hspace{\parindent} The morphology of the dust around the {\HII}
regions is displayed in Figures~\ref{fig:LMC_HII} and
\ref{fig:SMC_HII}.  There is much in the literature pertaining to
{\HII} region gas and dust morphologies
\citep[see][]{rela09,snid09,wats08,smit07,helo04}, and our results are
similar.  The warm dust at 24~$\mu$m is heavily peaked around the
centers of the {\HII} regions and near peaks in {\Ha} emission (see
Figures~\ref{fig:LMC_anc} and \ref{fig:SMC_anc}).  The 24~$\mu$m
emission drops off quickly, whereas, the aromatic 8~$\mu$m emission is
more extended.  The cold dust measured by 160~$\mu$m is very extended
and dominates the outer environs of the {\HII} region complexes.  On
smaller spatial scales than is resolved in this work, others have
noted that 8 $\mu$m is largely absent near the hot OB stars in the
center of {\HII} regions, possibly due to the destruction of aromatics
or low dust density near the inner OB stars
\citep{snid09,wats08,helo04}.  \citet{lebo07}, by observing NGC~3603,
provide spectroscopic evidence that aromatics are destroyed near the
central cluster, whereas, very small grains survive nearer the central
ionizing source.

\begin{figure*}
\includegraphics[angle=0, scale=0.43]{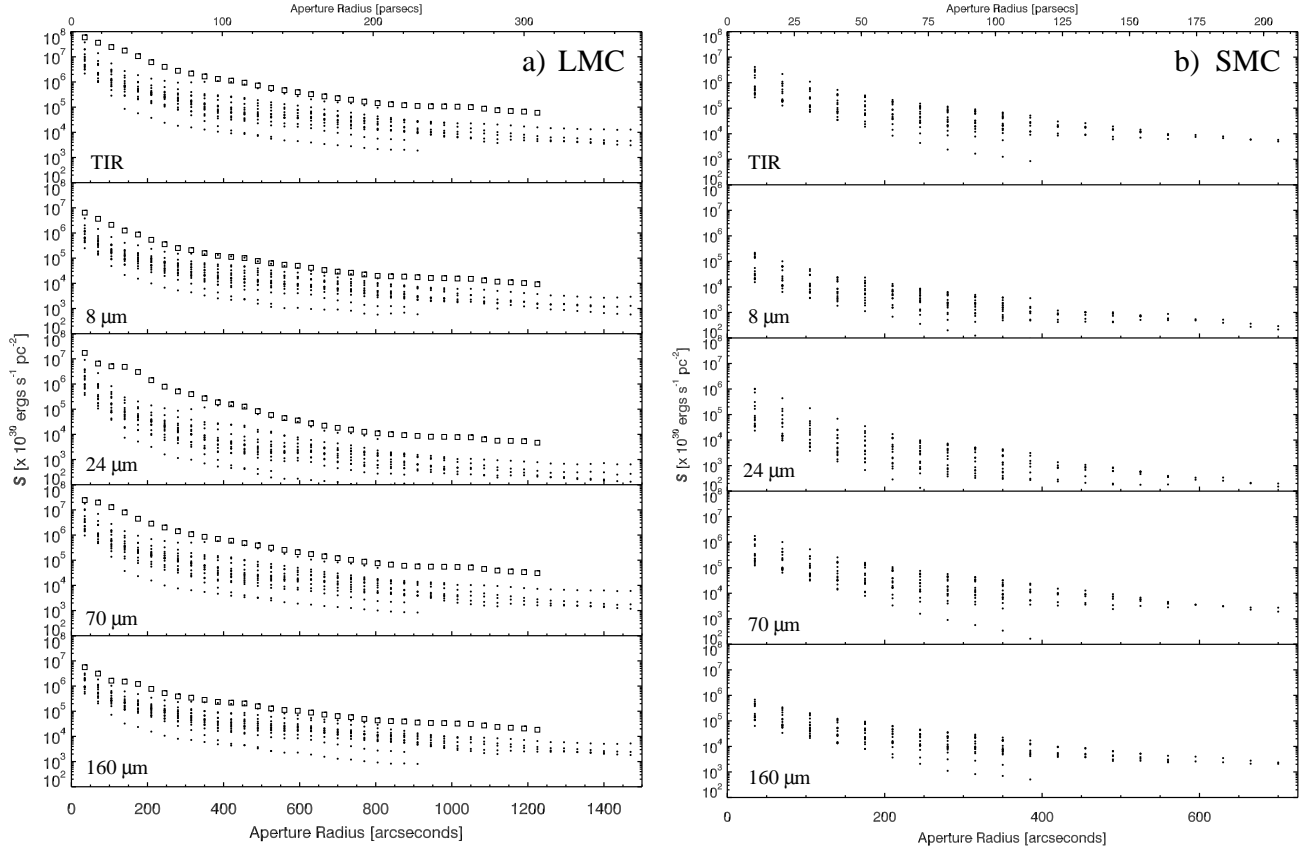}
\caption{Radial surface brightness profiles of the --a) 16 LMC {\HII}
  regions and the --b) 16 SMC {\HII} regions.  The surface
  brightnesses are in units of $10^{30}$~[ergs~s$^{-1}$~pc$^{-2}$],
  and the aperture radii are in units of [arcseconds] with spatial
  units of [parsecs] on the top axis.  The panels, from top to bottom,
  are TIR, 8~$\mu$m, 24~$\mu$m, 70 $\mu$m, and 160~$\mu$m.  The
  $\pm1\sigma$ statistical errors are on the order of the point sizes.
  The boxes in panel~(a) are the points associated with the measured
  surface brightness profile of 30~Doradus (N157).  The original LMC
  MIPS 70~$\mu$m data are from \citet{meix06}, and the original SMC
  MIPS 70~$\mu$m data are from {\gord}.}\label{fig:rad_SBR}
\end{figure*}

Within the limiting resolution of this work, the dust emission at
every wavelength peaks near the centers of the {\HII} regions, as we
show by plotting the LMC and SMC {\HII} region surface brightness
radial profiles in Figure~\ref{fig:rad_SBR}.  The TIR is the
bolometric IR flux calculated in section~\ref{SEC:TIR}, and the
$1\sigma$ uncertainties are on the order of the sizes of the points.
The scatter can be understood as being due to intrinsic differences in
the SEDs of the {\HII} regions.  30~Doradus (N157) has the highest
surface brightness profile at all wavelengths, as shown in
Figure~\ref{fig:rad_SBR}$a$ (boxes).  The surface brightnesses radial
profiles plotted here will be used to examine the broad {\HII} complex
sizes in section~\ref{size}.

The single-band IR surface brightness profiles in
Figure~\ref{fig:rad_SBR} are similar to the bolometric (TIR) surface
brightness profiles in that they all peak in the center of the {\HII}
regions, even for the coldest dust at 160~$\mu$m.  The central surface
brightness is lower for the SMC {\HII} regions than the LMC {\HII}
regions by about an order of magnitude, for all bands.  This is
possibly a resolution effect.  The resolution of our work, using our
smallest aperture, is $\sim20$ pc in diameter.  Any {\HII} region
smaller than this size, i.e., not resolved, will have an artificially
low surface brightness.  This might conceivably affect both the LMC
and SMC, but the SMC {\HII} luminosity function has a slightly steeper
slope and a lower high mass cutoff than the LMC \citep{kenn86}.  Thus,
the SMC may have a larger number of {\HII} regions that are small
enough to be unresolved.  To better study this, a larger sample of LMC
and SMC {\HII} region surface brightnesses, of similar luminosities,
will need to be analyzed.  Also, we need a better understanding of how
IR {\HII} region luminosity functions correlate with the {\Ha}
luminosity functions.

The 8~$\mu$m fluxes peak in the center despite studies showing a lack
of aromatics near the hot central OB stars and an abundance of
aromatics on the surface of outlying PDRs
\citep{rela09,snid09,wats08}.  Our inner aperture resolution of
$\sim20$ pc diameter is not as high as many of those previous studies.
We likely are not able to resolve the destruction or removal of the
aromatics near the central hot stars, and geometry may play a role via
foreground emission.  There is also the possible contamination in the
8~$\mu$m band of non-aromatics, such as {\HI}, [\ArII], and [\ArIII]
\citep[see][]{peet02,lebo07}.  It is unlikely that these other lines
dominate the emission at 8~$\mu$m, but they may contribute a
significant fraction.  This may be particularly true in the SMC where
there is a paucity of aromatics.

\section{DISCUSSION}
\label{disc}
\subsection{Normalized Radial SEDs}
\label{disc:seds}
\hspace{\parindent}From the results of section~\ref{res:SEDs}, we find
that most of the {\HII} regions in the LMC and SMC have IR SEDs that
peak around 70~$\mu$m.  Thus, we expect 70~$\mu$m to be a dominant
contributor to the bolometric IR flux.  However, there is intrinsic
scatter in the radial profiles of all of the IR bands (see
section~\ref{res:pro}).  To better understand the spatial variations
of the IR band emission in LMC and SMC {\HII} regions, we analyze the
radial SEDs at radii from near their core ($\sim10$ pc) to their
outskirts.  Because we are interested in how the IR dust bands relate
to the bolometric flux, we use the radial SED fluxes normalized by the
TIR, where the normalized flux at frequency $\nu$ is defined as
\begin{equation}
F_{\mbox{\tiny{N}}},_{\nu} = \nu~F_{\nu}/\mbox{TIR}.
\end{equation}

\begin{figure*}
\includegraphics[angle=0, scale=0.5]{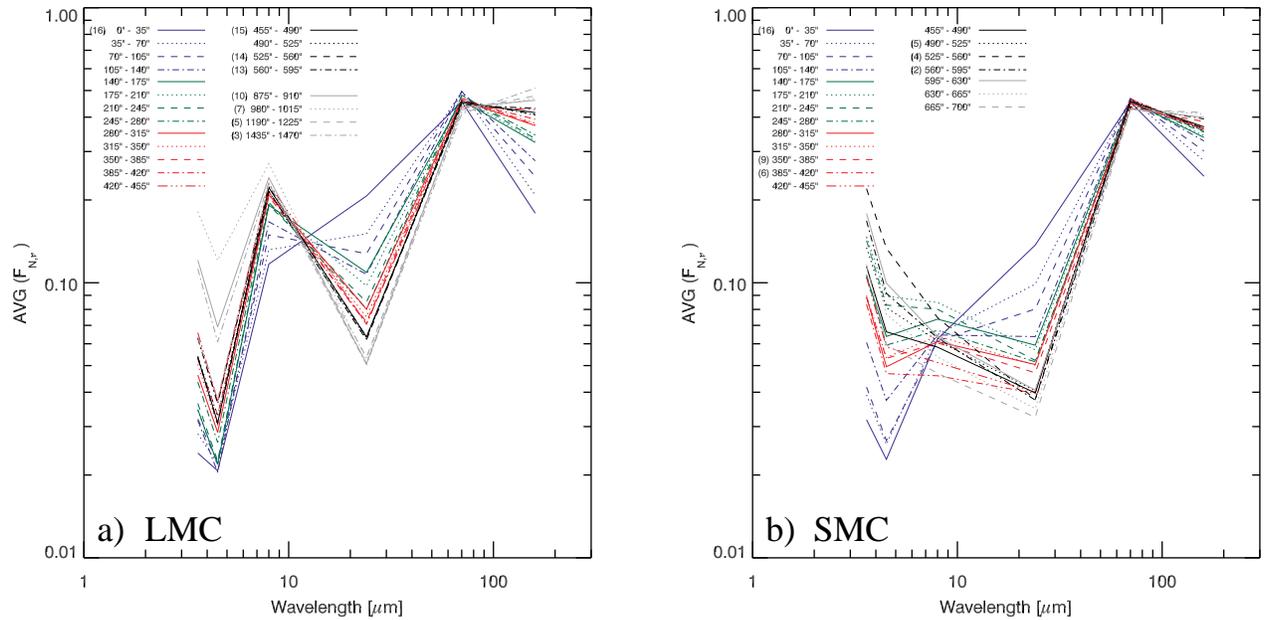}
\caption{Average normalized radial IR SEDs for the --a) LMC and --b)
  SMC.  Each line represents the average of the normalized (by TIR)
  fluxes of all of the {\HII} regions in an annulus with inner and
  outer radii (labeled in the upper left).  The parenthetical numbers
  to the left of the labels correspond to the number of {\HII} regions
  included for each annulus.  The colors of the lines go from the
  inner annuli (blue), to middle annuli (green and red), and finally
  to the outer annuli (black and gray).}\label{fig:SEDs_NORM}
\end{figure*}

In Figure~\ref{fig:SEDs_NORM}, we show the average normalized (by TIR)
flux, in units of $F_{\mbox{\tiny{N}}},_{\nu}$, for each annulus in
the LMC (panel a) and SMC (panel b).  Each line represents an average
of all of the {\HII} regions with a given annulus.  For both the LMC
and SMC, the inner annuli show relatively strong average
$F_{\mbox{\tiny{N}}},_{24}$ values that drop as the annuli proceed
radially outward.  The average $F_{\mbox{\tiny{N}}},_{70}$ values in
both the LMC and SMC remain relatively constant, with slight
variations.  The average $F_{\mbox{\tiny{N}}},_{160}$ values for both
the LMC and SMC behave opposite to that of the average
$F_{\mbox{\tiny{N}}},_{24}$ values in that they increases radially
outward from the center of the {\HII} region.

The normalized 8~$\mu$m aromatic emission behaves very differently in
the LMC than in the SMC.  In the LMC the average
$F_{\mbox{\tiny{N}}},_{8}$ values increase slightly with
radius.  In the SMC the average $F_{\mbox{\tiny{N}}},_{8}$
values do not follow an obvious trend with radius.  The 8~$\mu$m
emission in the SMC {\HII} regions on average accounts for about $7\%$
of their TIR, while in the LMC {\HII} regions, aromatics account for
$15\%$ of the TIR.  The lower normalized 8~$\mu$m values in the SMC
are at least partially due to the lower intrinsic fluxes at 8~$\mu$m
for SMC {\HII} regions, and the SMC in general (see
Figure~\ref{fig:SEDs_FULL}).  The lack of a radial trend for the
8~$\mu$m emission in the SMC may be a combination of the lack of
aromatics, in which we would expect a trend similar to the LMC
8~$\mu$m where aromatics emission increases away from the central
ionizing source, and potential contributions to the 8~$\mu$m from
other bands.  For instance, the [\ArII] nebular emission line is not
expected to contribute much to the 8~$\mu$m emission \citep{lebo07},
but the lack of strong aromatics in the SMC means that this weaker
line, along with {\HI} and [\ArIII] emission \citep{peet02,lebo07},
may be more significant factors in the observed 8~$\mu$m fluxes.  The
contribution of non-aromatic emission to the 8~$\mu$m fluxes within
SMC {\HII} regions needs to be more fully explored.  The lower SMC
8~$\mu$m emission is further discussed in section~\ref{radial}.

The $F_{\mbox{\tiny{N}}},_{70}$ values are nearly constant
among all annuli for both the LMC and SMC ($70~\mu\mbox{m} \approx
45\%$ TIR).  \textit{Thus, the SEDs in each annuli must all peak near
70~$\mu$m and the peak of the SED must not considerably vary as we
spatially observe the {\HII} regions from their centers outward}.  We
further explore the implications of this result in the following
sections.

\subsection{Physical Sizes of {\HII} Region Complexes}
\label{size}
\hspace{\parindent}The appropriate size aperture to use for
extragalactic studies of {\HII} regions is unknown and is typically
set by the limiting resolution of the instrument because many of the
{\HII} regions are not resolved in the IR at extragalactic distances.
An incorrect aperture size will add uncertainties to any SFR
calculation because the accumulated flux may not all be from the
{\HII} region of interest.  We explore the {\HII} complex size scales,
as probed by the 8, 24, 70, and 160~$\mu$m dust emission, and compare
them to the size scales found using the TIR flux.

We quantitatively probe the sizes of the LMC and SMC {\HII} regions by
tracing the TIR, 8, 24, 70, and 160~$\mu$m radial profiles (see
Figure~\ref{fig:rad_SBR}) and calculate the radius at which each
band's surface brightness drops by $95\%$ ($\sim2\sigma$) from that
measured in the inner $35\arcsec$ ($\sim10$ pc) radius
aperture\footnote{We choose to use a common cutoff for all of the
bands.  A $3\sigma$ cutoff was attempted, however, it was too strict
for many of the {\HII} regions.  The surface brightnesses rarely
dropped to below $3\sigma$ of the central aperture for the coldest
dust at 160~$\mu$m.  This is due to the relatively shallow slopes of
the radial 160~$\mu$m surface brightnesses}.  All four dust bands peak
in the center aperture for every {\HII} region (see
Figure~\ref{fig:rad_SBR}), and the central aperture is where we are
most certain that young stars dominate the UV output.

The results of our size determinations can be found in
Table~\ref{tab:HII_sizes}.  The size for each LMC and SMC {\HII}
region complex is listed along with the radius in arcseconds where the
surface brightness drops to less than $5\%$ of the surface brightness
of the inner aperture.  The mean, standard deviation, and fractional
error (standard deviation divided by the mean) for the LMC and SMC
{\HII} complexes are also tabulated.  The {\HII} region sizes have
large standard deviations about the mean values.  The errors, as a
fraction of the mean, are $31\%-39\%$ for all bands.

\begin{deluxetable}{lrrrrr}
\tabletypesize{\scriptsize}
\tablecolumns{6}
\tablewidth{0pt}
\tablecaption{Sizes of {\HII} Region Complexes\label{tab:HII_sizes}}
\tablehead{
\colhead{}                  &
\colhead{TIR}               &
\colhead{$8 \mu$m}          &
\colhead{$24 \mu$m}         &
\colhead{$70 \mu$m}         &
\colhead{$160 \mu$m}        \\
\colhead{Name}              &
\multicolumn{5}{c}{Radius (arcsec)}}
\startdata
\multicolumn{6}{c}{LMC} \\ 
\hline
N4                   & 140           & 210          & 105          & 140         & 210             \\
N11                  & 245           & 490          & 210          & 245         & 420             \\ 
N30                  & 210           & 280          & 140          & 210         & 350             \\ 
N44                  & 315           & 385          & 210          & 315         & 350             \\ 
N48                  & 315           & 385          & 140          & 315         & 420             \\ 
N55                  & 245           & 245          & 210          & 245         & 245             \\ 
N59                  & 175           & 245          & 140          & 175         & 245             \\ 
N79                  & 140           & 175          & 105          & 140         & 210             \\ 
N105                 & 175           & 210          & 140          & 175         & 245             \\ 
N119                 & 385           & 490          & 280          & 420         & 490             \\ 
N144                 & 210           & 280          & 175          & 210         & 315             \\ 
N157                 & NA$^a$        & 280          & NA$^a$       & 315         & NA$^a$          \\ 
N160                 & 175           & 245          & 140          & 210         & 490             \\ 
N180                 & 315           & 490          & 245          & 315         & 490             \\ 
N191                 & 140           & 140          & 105          & 140         & 140             \\ 
N206                 & 315           & 350          & 210          & 315         & 385             \\ 
\hline
Mean                 & 233           & 306          & 170          & 243         & 334             \\
Std dev              & 79            & 113          & 54           & 82          & 114             \\
Fr.error$^c$         & 0.34          & 0.37         & 0.32         & 0.34        & 0.34            \\
\hline
\multicolumn{6}{c}{SMC} \\ 
\hline
DEM74                & NA$^b$        & NA$^b$       & 245          & NA$^b$      & NA$^b$          \\  
N13                  & 245           & 210          & 140          & 245         & 350             \\  
N17                  & 280           & 280          & 210          & 280         & 315             \\  
N19                  & 280           & 280          & 245          & 280         & 315             \\  
N22                  & 245           & 245          & 140          & 245         & 315             \\  
N36                  & 455           & 385          & 385          & 455         & 490             \\  
N50                  & 525           & 490          & 315          & 525         & 560             \\  
N51                  & 280           & 245          & 140          & 280         & 350             \\  
N63                  & 210           & 245          & 210          & 210         & 210             \\  
N66                  & 210           & 210          & 175          & 245         & 280             \\  
N71                  & 140           & 280          & 140          & 140         & 210             \\  
N76                  & 315           & NA$^b$       & 245          & 315         & NA$^b$          \\  
N78                  & 140           & 140          & 105          & 140         & 210             \\  
N80                  & 350           & 280          & 210          & 315         & 385             \\  
N84                  & 315           & 280          & 210          & 315         & 350             \\  
N90                  & 175           & 210          & 140          & 175         & 210             \\
\hline
Mean                 & 278           & 270          & 203          & 278         & 325             \\
Std dev              & 107           & 84           & 74           & 105         & 105             \\
Fr.error$^c$        & 0.39          & 0.31         & 0.36         & 0.38        & 0.32              
\enddata
\tablecomments{{\HII} complex sizes are computed using the method described in section~\ref{size}.}
\tablenotetext{\mbox{a}}{Not calculated due to flux saturation in the core of the {\HII} region.}
\tablenotetext{\mbox{b}}{Not calculated due to a shallow radial profile; the maximum {\HII} region aperture is too small.}
\tablenotetext{\mbox{c}}{$\mbox{Fr.error} = \mbox{Std dev} / \mbox{Mean}$.}
\end{deluxetable}

For the TIR in the LMC {\HII} regions, the minimum/maximum radii
measured are 140/385 arcsec (35/97 pc), respectively.  For the TIR in
the SMC {\HII} regions, the minimum/maximum radii measured are 140/525
arcsec (41/154 pc), respectively.  The computed sizes listed in
Table~\ref{tab:HII_sizes} are not drastically different for the LMC
and the SMC, particularly given the large standard deviations.

\begin{figure*}
\includegraphics[angle=0, scale=0.68]{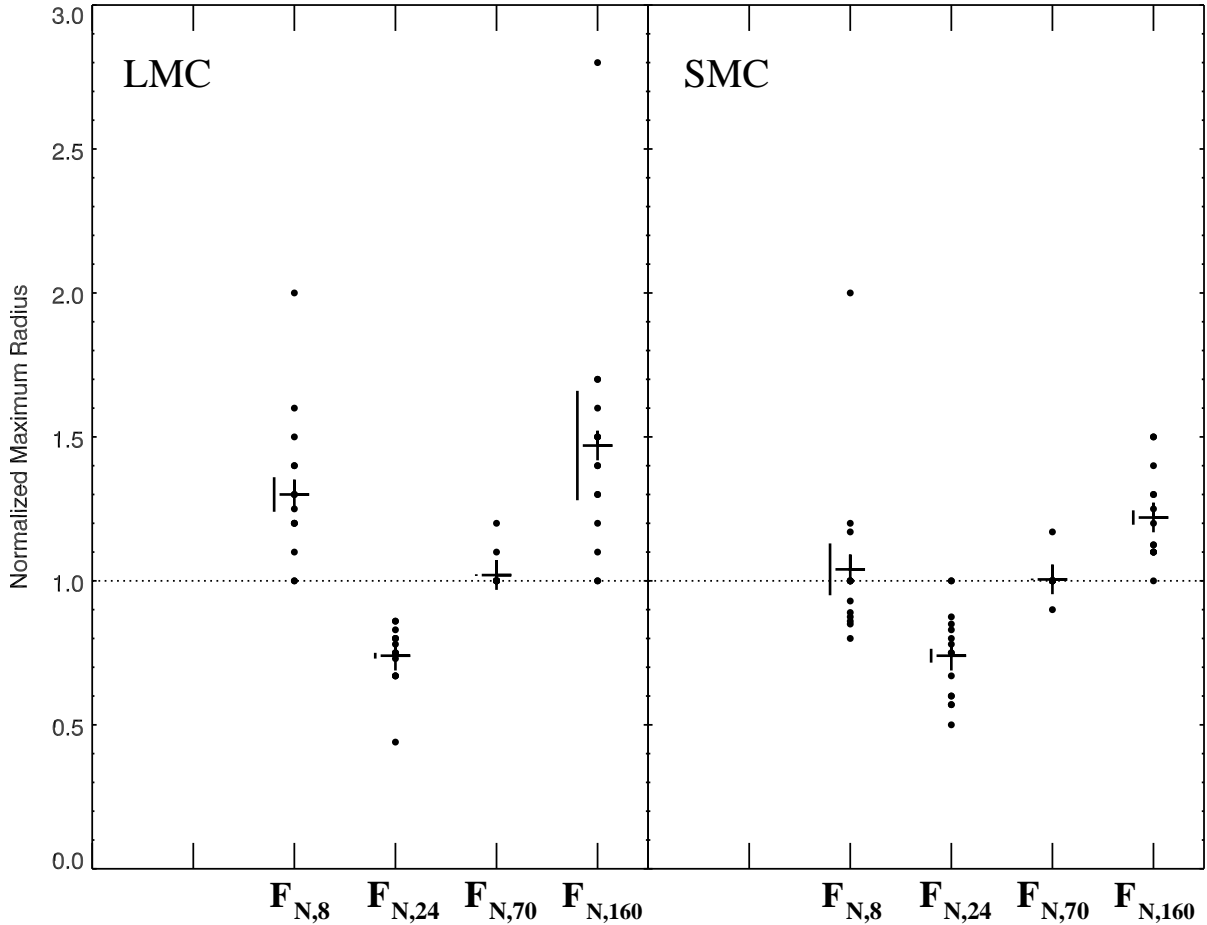}
\caption{LMC and SMC {\HII} region sizes of the 8, 24, 70, and
  160~$\mu$m fluxes normalized by the sizes computed using TIR.  The
  LMC is plotted in the left panel, and the SMC is plotted in the
  right panel.  The points are the data, the crosses are the mean
  values, and the solid lines to the left of the data are the
  $\pm1\sigma$ standard deviations.  All of the {\HII} regions in
  Table~\ref{tab:HII_sizes} are included, but many of the individual
  {\HII} regions overlap due to discrete $35\arcsec$ increasing radii
  for successive annuli.  The horizontal dotted line marks the
  positions where the sizes of the monochromatic bands are equal to
  the computed sizes measured using TIR.}\label{fig:HII_Sizes}
\end{figure*}

The sizes calculated from TIR, though not identical, are close to the
sizes calculated using the 70~$\mu$m emission.  This can most clearly
be seen in Figure~\ref{fig:HII_Sizes} where we plot the {\HII} region
8, 24, 70, and 160~$\mu$m sizes, normalized by the TIR sizes.  The
mean $F_{\mbox{\tiny{N}}},_{70}$ size falls nearly on the vertical
dotted line associated with TIR for both the LMC and SMC.  The
standard deviation in the sizes measured using
$F_{\mbox{\tiny{N}}},_{70}$ is smaller than for the sizes measured
using the other normalized bands.  This is a another consequence of
70~$\mu$m being near the peak of dust SED (see
Section~\ref{disc:seds}).  The mean size of the {\HII} regions using
either TIR or 70~$\mu$m is $60\pm20$ pc the LMC and $80\pm30$ pc for
the SMC.

The calculated 8~$\mu$m sizes in the LMC are as large as or larger
than sizes measured using 70~$\mu$m or TIR, with the exception of
30~Doradus (N157).  There is no such trend in the SMC where the sizes
traced by the 8~$\mu$m emission may be larger or smaller than the
sizes calculated using the 70~$\mu$m dust emission.  This is shown in
Figure~\ref{fig:HII_Sizes} where the sizes of the {\HII} regions, as
measured by $F_{\mbox{\tiny{N}}},_{8}$, are systematically larger
than TIR in the LMC but roughly scattered around TIR in the SMC.  This
trend is consistent with the 8~$\mu$m flux following a shallower
decline with radius in the LMC relative to the SMC.  The mean size of
the {\HII} regions using 8~$\mu$m is $75\pm35$ pc for the LMC and
$80\pm25$ pc for the SMC.

All of the {\HII} regions are smallest when calculated using the
24~$\mu$m emission.  This is consistent with others who have found the
warm small dust grains probed by 24~$\mu$m being more heavily peaked
near the center of star forming regions \citep[e.g.,][]{rela09}.  None
of the {\HII} regions have sizes measured via
$F_{\mbox{\tiny{N}}},_{24}$ in Figure~\ref{fig:HII_Sizes} that are
larger than the sizes calculated via TIR.  The mean size of the {\HII}
regions using 24 $\mu$m is $40\pm15$ pc for the LMC and $60\pm20$ pc
for the SMC.

The mean size of the {\HII} regions using 160~$\mu$m is $85\pm30$ pc
for the LMC and $95\pm30$ pc for the SMC.  We have shown in
Figure~\ref{fig:SEDs_NORM} that the $F_{\mbox{\tiny{N}}},_{160}$
averages in the {\HII} region SEDs are largest at large radii (also
discussed in section~\ref{radial}).  Consequently, the {\HII} region
sizes, as probed by 160~$\mu$m emission, are generally larger than the
sizes calculated using the other IR dust bands.  When the sizes
measured via the TIR and 160~$\mu$m emission are compared (see
Figure~\ref{fig:HII_Sizes}), the sizes calculated using the cold
160~$\mu$m emitting dust are systematically larger or the same size as
the TIR in both the LMC and SMC.  The sizes measured by the cold dust
emission at 160~$\mu$m are the largest despite the fact that the
160~$\mu$m surface brightnesses peak in the center of the {\HII}
regions (see Figure~\ref{fig:rad_SBR}).  The increasing importance of
the 160~$\mu$m emission to the TIR in the outer radii of the {\HII}
regions causes a relatively shallow decline in flux with radius.  The
computed large {\HII} region cold dust sizes are a consequence of the
pervasiveness of the cold dust emission to large radii.

From this analysis, we conclude that the sizes of {\HII} regions, as
probed by dust, depend greatly on the wavelength observed but also
depend greatly on the individual {\HII} region due to the large
deviations in each band.  The warmer dust probed by 24~$\mu$m will
give smaller {\HII} region sizes than the cold dust probed by 160
$\mu$m.  The sizes probed by 70~$\mu$m emission is nearly identical to
the sizes probed by the TIR, with little scatter.  This is a
consequence of the IR SEDs peaking near 70~$\mu$m at all spatial
distances for nearly every {\HII} region (see
section~\ref{disc:seds}).  Thus, when taking into account the limits
of resolution and aperture size for extragalactic studies, 70~$\mu$m
is the most ideal of the monochromatic IR star formation indicators
because it will trace the TIR spatially and extend to physically
larger distances than, say, the 24~$\mu$m emission.  Furthermore, we
show in Figure~\ref{fig:HII_Sizes}$a$ that for most of the {\HII}
regions in our sample, an aperture radius of $\sim100$ pc would be
adequate for acquiring most of the 70~$\mu$m flux.

The size distribution of the nebular portion of {\HII} regions, as
traced by {\Ha}, has been extensively studied and argued to follow
either a power law \citep{kenn80} or an exponential law \citep{van81}.
Given that the luminosity function is calculated to follow a power
law, \citet{oey03} argue that the size distribution must also follow a
power law because the luminosity measurements come from ionizing
photons within a specific nebular volume.  For the Magellanic Clouds,
\citet{kenn86} use an exponential function to estimate the scale
length for {\HII} region sizes to be $\sim80$ pc.  From our data, we
cannot determine a statistically significant size distribution for the
LMC or SMC because our sample is too small.  The initial selection of
our 32 {\HII} regions was by eye which will bias any size
distributions created from this data.  A more objective IR selection
of a larger number of LMC and SMC {\HII} regions is required for an
analysis of the LMC and SMC {\HII} region size distributions,
particularly at the small end.  Ignoring the warmer dust probed via
24~$\mu$m, the sizes we do measure are roughly the same as that of the
\citet{kenn86} derived {\Ha} size distribution scale length.

It must be stated that the sizes we measure include the general star
forming regions.  The dust emission is high out to relatively large
distances.  Our size estimates are not directly comparable to the
{\Ha}-derived nebular sizes.  Although there may be similarities in
that the IR-measured sizes of the LMC {\HII} regions should be
somewhat larger than those in the SMC because the {\Ha} and
\textit{IRAS} 100~$\mu$m luminosity functions in the LMC have a higher
mass cut-off than in the SMC \citep[see discussion in
Section~\ref{res:SEDs};][]{kenn86,liva07}.  The large standard
deviations of the IR-derived sizes is real because the scatter in the
surface brightness profiles from which they are derived is real (see
section~\ref{res:pro}).  This scatter is probably introduced by
sampling {\HII} regions of different high mass stellar populations and
evolutionary histories, which are two components that shape the {\Ha}
luminosity function \citep{oey98}.

\subsection{TIR Normalized Radial Profiles}
\label{radial}
\begin{figure*}
\includegraphics[angle=0, scale=0.23]{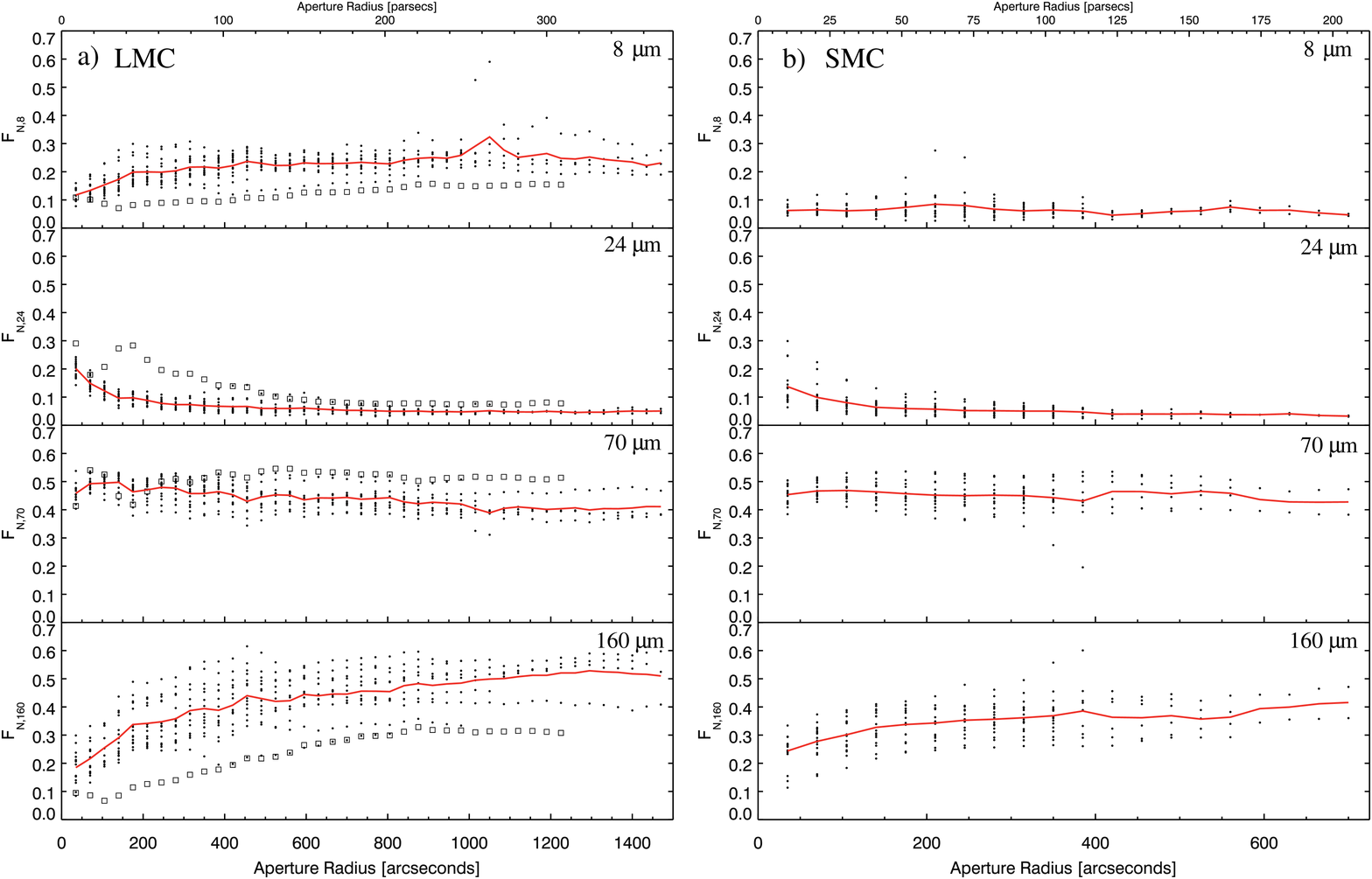}
\caption{Radial profiles, from top to bottom, of the 8, 24, 70, and
160~$\mu$m fluxes, normalized by TIR, for the --a) 16 LMC {\HII}
regions and --b) 16 SMC {\HII} regions.  The $\pm1\sigma$ statistical
errors are on the order of the point sizes.  The red line traces the
average normalized flux, calculated for each annulus.  Spatial units
(parsecs) are labeled on the top axis.  The boxes in panel (a)
represent 30~Doradus (N157) and are not included in the average value
(red line) because 30~Doradus is saturated around the core in 24 and
160~$\mu$m.}\label{fig:rad_NORM}
\end{figure*}

\hspace{\parindent} Plotted in Figure~\ref{fig:rad_NORM} are the 8,
24, 70, and 160~$\mu$m radial profiles, normalized by TIR,
($F_{\mbox{\tiny{N}}},_{\nu}$), of the LMC and SMC {\HII} regions.
For the LMC {\HII} regions, the $F_{\mbox{\tiny{N}}},_{8}$ emission
is truncated near the center of the {\HII} regions (down to $\sim10\%$
TIR) and is relatively constant ($\sim20\% - 25\%$ TIR) for annuli
with radii greater than $\sim 200\arcsec$ ($\sim 50$ pc).  The
$F_{\mbox{\tiny{N}}},_{24}$ emission is sharply peaked near the
centers of the {\HII} regions ($\sim 20\%$ TIR) and falls to a modest
level ($\sim 5\% - 10\%$ TIR) for annuli with radii greater than $\sim
200\arcsec$ ($\sim 50$ pc).  The $F_{\mbox{\tiny{N}}},_{70}$ emission
is relatively constant for all radii probed here ($\sim 40\% - 50\%$
TIR).  The $F_{\mbox{\tiny{N}}},_{160}$ emission increases
dramatically with aperture/annuli radii, from ($\sim 20\%$ TIR) near
the core to ($\sim 50\%$ TIR) near the outer annuli.  The scatter
about the mean for $F_{\mbox{\tiny{N}}},_{160}$ is larger than for
the other emission bands.  The 8~$\mu$m outliers in
Figure~\ref{fig:rad_NORM}$a$ are from the N4 {\HII} region, where two
bright 8~$\mu$m sources are included in the outer annuli at $\sim260$
pc from the center (see N4 in Figure~\ref{fig:LMC_HII}).

For the SMC {\HII} regions, the $F_{\mbox{\tiny{N}}},_{8}$ does
not show an increase from the central core as it does in the LMC.  The
$F_{\mbox{\tiny{N}}},_{8}$ is consistently around $5\%$ TIR at
all radii, which is significantly smaller than in the LMC.  The
$F_{\mbox{\tiny{N}}},_{24}$ emission is peaked in the center as
it is for the LMC, although to a smaller magnitude ($\sim15\%$ TIR).
The $F_{\mbox{\tiny{N}}},_{70}$ emission in the SMC is
consistent with that for the LMC, at all annuli radii.  The
$F_{\mbox{\tiny{N}}},_{160}$ emission is also similar to that
for the LMC.

Due to the change in the shape of the
$F_{\mbox{\tiny{N}}},_{24}$ profile at low radii, 24~$\mu$m may
not be as robust a tracer of TIR (or obscured SFR) as 70~$\mu$m, which
has a relatively constant profile with radius.  An observer would need
to develop a conversion from 24~$\mu$m to TIR that is dependent on the
radius of the aperture used for the measurement.

The average SMC $F_{\mbox{\tiny{N}}},_{8}$ profile in
Figure~\ref{fig:rad_NORM} is depressed relative to the LMC at all
scales.  The same is true, although to a lesser extent, for the
$F_{\mbox{\tiny{N}}},_{24}$ profile.  The 8 and 24~$\mu$m fluxes are
at least partially associated with aromatics/small grains undergoing
stochastic heating \citep{puge89,drai07}.  The low values of
$F_{\mbox{\tiny{N}}},_{8}$ and $F_{\mbox{\tiny{N}}},_{24}$ in the
SMC relative to the LMC are an indication that the SMC has fewer of
these aromatics/small grains.  This is consistent with the results in
{\gord} who find that the SMC has an aromatic fraction of $1.1\%$,
which puts the aromatic abundance in the SMC on the low end of the
\citet{drai07b} \textit{Spitzer} SINGS sample.  This is also
consistent with the SMC's low metallicity and high radiation field
hardness \citep{gord08}.  The strong dependence of the host galaxy's
aromatic/small grain population presents a problem for using either of
these bands as a monochromatic star formation indicator.  Previous
works have noted some of these potential problems when using 8 or
24~$\mu$m as star formation indicators
\citep[e.g.,][]{dale05,calz05,calz07}.

\begin{figure*}
\includegraphics[angle=0, scale=0.215]{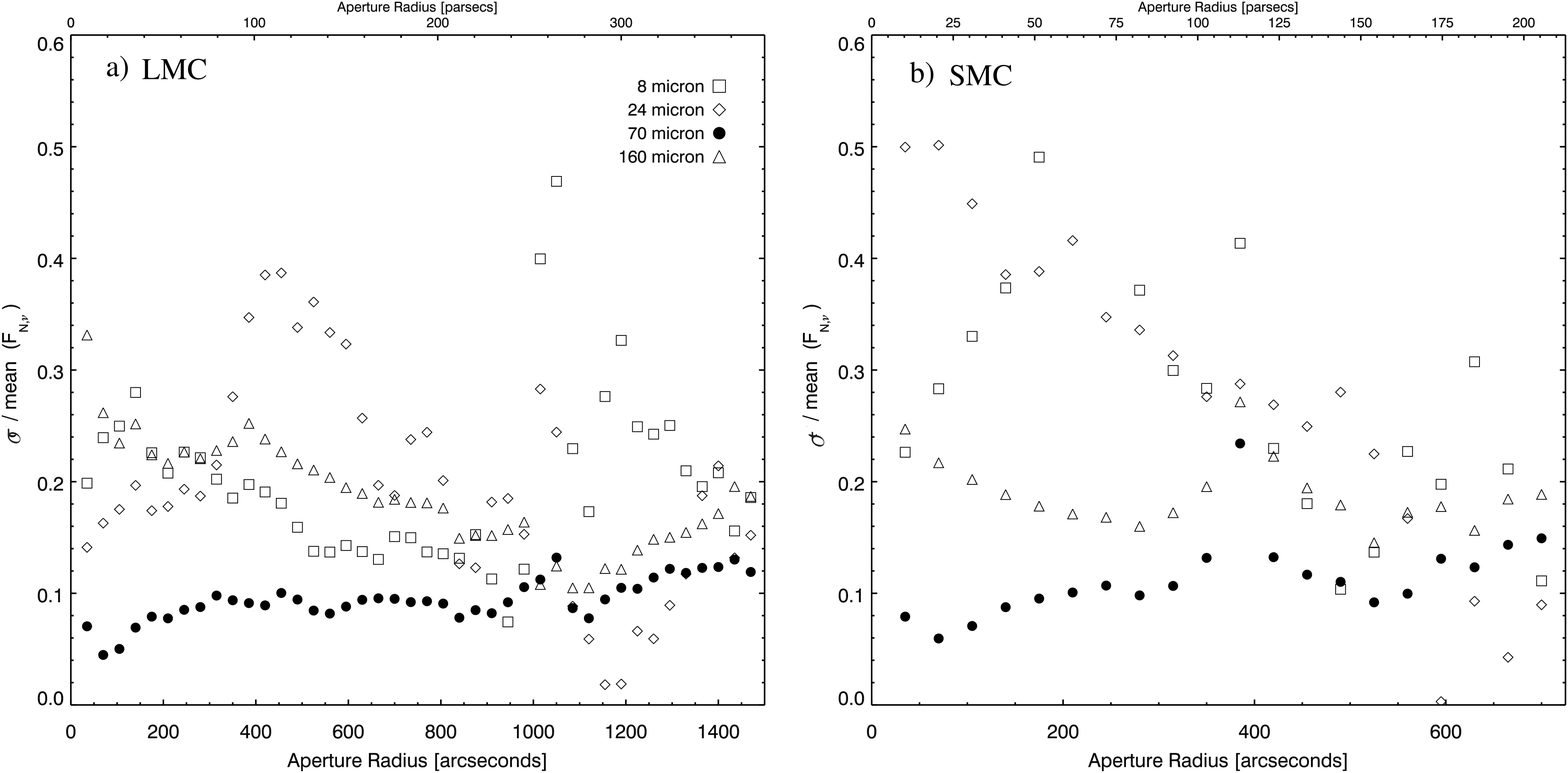}
\caption{Fractional errors ($\sigma$/mean) of the 8 (open squares), 24
  (open diamonds), 70 (filled circles), and 160~$\mu$m (open
  triangles) TIR normalized radial profiles for the --a) LMC and --b)
  SMC.  Physical spatial units (parsecs) are given on the top axis.
  Due to saturation in 24 and 160~$\mu$m, the 30~Doradus (N157) {\HII}
  region is not included in the LMC computations of fractional
  errors}\label{fig:rad_NORM_sigma}
\end{figure*}

In Figure~\ref{fig:rad_NORM_sigma}, we show the fractional error
($1\sigma/\mbox{mean}$) of the $F_{\mbox{\tiny{N}}},_{\nu}$
measurements for each annulus.  Plotted are the fractional errors with
respect to annuli radius for the LMC (panel a) and SMC (panel b).  For
all radii, except for perhaps at the largest radii measured here where
low number statistics dominate, the fractional error of the
$F_{\mbox{\tiny{N}}},_{70}$ indicator is lower than the other IR dust
emission bands and is $\sim5\% - 12\%$ the mean value at radii less
than 240 pc in the LMC and SMC.  The fractional error of
$F_{\mbox{\tiny{N}}},_{24}$ is $>0.15$ until the largest radii where
low number statistics become an issue.

Using the IR fluxes, normalized by the TIR, we find that the 70~$\mu$m
emission is a robust tracer of the TIR, and thus the star formation
obscured by dust.  This is because the normalized 70~$\mu$m flux is
relatively constant (70~$\mu\mbox{m}\sim0.40-0.50~\mbox{TIR}$) for all
radii measured, does not systematically vary between the LMC and SMC,
and has the smallest fractional errors at nearly every radii.  These
results are consequences of the fact that the SEDs of the {\HII}
region annuli in our sample all peak near $70~\mu$m (see
section~\ref{disc:seds}).

These results will be less robust for galaxies that have {\HII}
regions that are hotter than the typical LMC or SMC {\HII} region.
However, we do not expect this to be the case because the hottest
{\HII} region in our sample, 30~Doradus, would fall on the high end of
any {\HII} region {\Ha}-derived luminosity function \citep{kenn89}.
If hotter {\HII} regions are more common in a particular galaxy, they
will have dust profiles more like 30~Doradus (see boxes in
Figure~\ref{fig:rad_NORM}$a$) where the 24~$\mu$m flux is a larger
percentage of the TIR out to larger radii.  Consequently, the
normalized 70~$\mu$m flux is reduced near the core.  This produces an
increasing slope in $F_{\mbox{\tiny{N}}},_{70}$ until some
radius where the 24 $\mu$m emission is no longer a dominant
contributor to the TIR.  It is probable that we would see this effect
on smaller scales than is probed in this work for all {\HII} regions.

The photons of hotter {\HII} regions will also affect the observed 8
and 160~$\mu$m normalized fluxes.  The hotter {\HII} regions may push
the photoionization front to larger radii causing a decrement of
aromatics near the core.  Thus, the peak normalized 8~$\mu$m flux
would occur at larger radii.  The 160~$\mu$m flux would not dominate
the TIR until even larger radii than is measured in this work.  If
hotter {\HII} regions are more common in other galaxies, then there
will be a larger scatter about the average
$F_{\mbox{\tiny{N}}},_{70}$ ratio with larger variations near
the core.

\subsection{Star Formation Rates}\label{sec:SFR}

\subsubsection{Radial SFR Densities}\label{sec_SFRrad}
\hspace{\parindent} For the star formation obscured by dust, we use
the TIR SFR computation from Equation (4) of \citet{kenn98}.  The
\citet{kenn98} obscured SFR equation is valid for starbursts, but dust
emission in {\HII} regions and starburst galaxies of varying
metallicities and ionization measures behave similarly
\citep{gord08,enge08}.  Dust obscured star formation in {\HII} regions
is similar to star formation in starbursts because starbursts are
usually very dusty and, furthermore, were likely the birthplaces of
most of the present-day stars \citep{elba03,sala09}.  The
\citet{kenn98} obscured SFR equation takes the form of
\begin{equation}\label{EQ:SFR}
\mbox{SFR}=4.5\times10^{-44}~L_{TIR},
\end{equation}
where SFR is the star formation rate in units of solar masses per
year.  The value $4.5\times10^{-44}$ is a constant derived from
spectral synthesis models, assumptions on the shape of the initial
mass function (IMF), and assumptions on the timescale of the star
formation event \citep{kenn98}.  $L_{TIR}$ is the TIR luminosity in
units of ergs per second.  Similarly, we use Equations~1 and 2 in
\citet{kenn98} to calculate the unobscured SFRs derived from UV and
{\Ha}, which rely on similar assumptions.  In all three cases a
Salpeter IMF is assumed \citep{salp55}.

It is noted by \citet{rela09} and \citet{helo04} that {\Ha} emission
peaks near the maximum 24~$\mu$m emission.  However, the peak UV
emission is offset from {\Ha} and 24~$\mu$m, presumably because the UV
emission is absorbed by the dust where the {\Ha} and 24~$\mu$m emit
the strongest \citep{rela09}.  We also find that {\Ha} peaks near the
peak 24~$\mu$m emission (see Figures~\ref{fig:LMC_anc} and
\ref{fig:SMC_anc}).  The UV fluxes in the LMC data indicate a peak UV
emission near the peak {\Ha} and 24~$\mu$m fluxes, but as a result of
our spatial resolution we cannot determine if the UV is offset from
the peak {\Ha} or 24~$\mu$m emission as is claimed by \citet{rela09}.
There are regions in the LMC with strong UV fluxes but weak or absent
{\Ha} and 24~$\mu$m fluxes.  Many of these UV fluxes are found in the
centers of noted {\HI} supergiant shells where hot OB stars and
supernovae remnants have cleared out much of the gas and dust and
triggered star formation on their peripheries
\citep[see][]{kim99,cohe03}.

\begin{figure*}
\includegraphics[angle=0, scale=0.42]{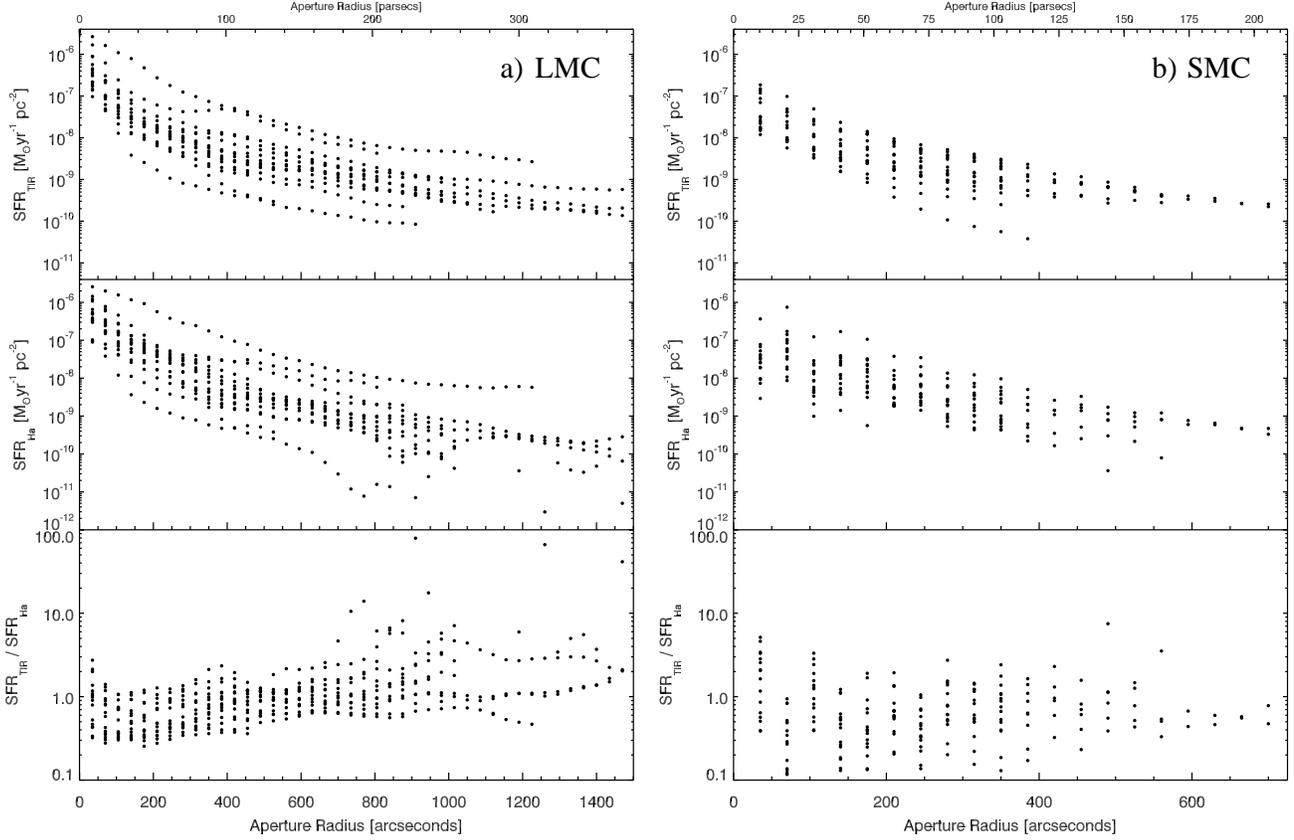}
\caption{Computed radial obscured (TIR--top panels) and unobscured
({\Ha}--middle panels) SFR densities ($M_{\odot}$ yr$^{-1}$ pc$^{-2}$)
for each of the --a) LMC and --b) SMC {\HII} regions, using the SFR
equations~2 and 4 from \citet{kenn98}.  The bottom panels are the
ratio of the obscured TIR-derived SFRs to the unobscured \Ha-derived
SFRs.  The radial units are in arcseconds with spatial radial units of
parsecs on the top axis.}\label{fig:SFR_Ha}
\end{figure*}

In Figure~\ref{fig:SFR_Ha}, we show the radial dependence of SFR
densities ($M_{\odot}$ yr$^{-1}$ pc$^{-2}$) derived using the obscured
SFR indicator (TIR) and the unobscured SFR indicator ({\Ha}) for the
LMC (panel a) and the SMC (panel b).  The magnitudes of the SFR
densities depend on the aforementioned assumptions about the IMF, star
formation timescales, and spectral synthesis models used
\citep{kenn98}.  However, the magnitudes are not crucial for this
analysis.  It is only the dependence with radius that we wish to
convey.  The general trends of the obscured and unobscured SFRs with
radius will be constant because all conversions from a star formation
indicator to a SFR depend upon a constant value.

The obscured and unobscured SFR densities in the LMC are larger than
in the SMC, which is expected given the larger LMC {\HII} region
luminosities (see Section~\ref{res:SEDs}) and the high-end cutoff at
larger luminosities for the LMC {\Ha} and IR luminosity functions
\citep{kenn86,liva07}.

For the obscured SFR densities, both the LMC and SMC show a peak in
the inner annuli and a decrease outward.  This is expected given that
the TIR emission peaks in the centers of the {\HII} regions as plotted
in Figure~\ref{fig:rad_SBR}.  The unobscured SFR densities, calculated
via {\Ha}, also peak in the center for LMC {\HII} regions.  Because
the 24~$\mu$m emission peaks where the TIR emission peaks in
Figure~\ref{fig:rad_SBR}, these results are further evidence, along
with the results in Figure~\ref{fig:LMC_anc}, that the 24~$\mu$m
emission traces the {\Ha} emission.  Along the same lines, the 8, 70,
and 160~$\mu$m peak fluxes will also trace the peak {\Ha} fluxes.
These results should break down on smaller scales than measured here
because the dust grains may be destroyed when near the central
ionizing stars.  For example, it is noted that the aromatics probed
with 8~$\mu$m are not observed to be any nearer the central OB stars
than in the outlying PDRs \citep{rela09,snid09,wats08,helo04}.

The {\Ha}-derived unobscured SFR densities are slightly offset from
the TIR-derived obscured SFR densities in the SMC (see middle panel of
Figure~\ref{fig:SFR_Ha}).  There is a slight decline of the {\Ha}
determined unobscured SFR densities in the inner aperture.  This may
well be a consequence of our limited {\Ha} resolution.  We aim to
explore this in the future with higher resolution data.

As shown in the bottom panel of Figure~\ref{fig:SFR_Ha}$a$, the ratios
of obscured SFR densities to the unobscured SFR densities slightly
increase with radius in the LMC.  If this trend is real then it
indicates that {\Ha} is more strongly peaked in the center of an
{\HII} region than TIR.  One possible explanation is that the TIR is a
measurement that includes dust at many temperatures.  As shown
earlier, the colder dust dominates at larger radii.  This trend is not
seen in the SMC, although, we do not trace many of the SMC {\HII}
regions out to large radii.

\begin{figure}
\includegraphics[angle=0, scale=0.42]{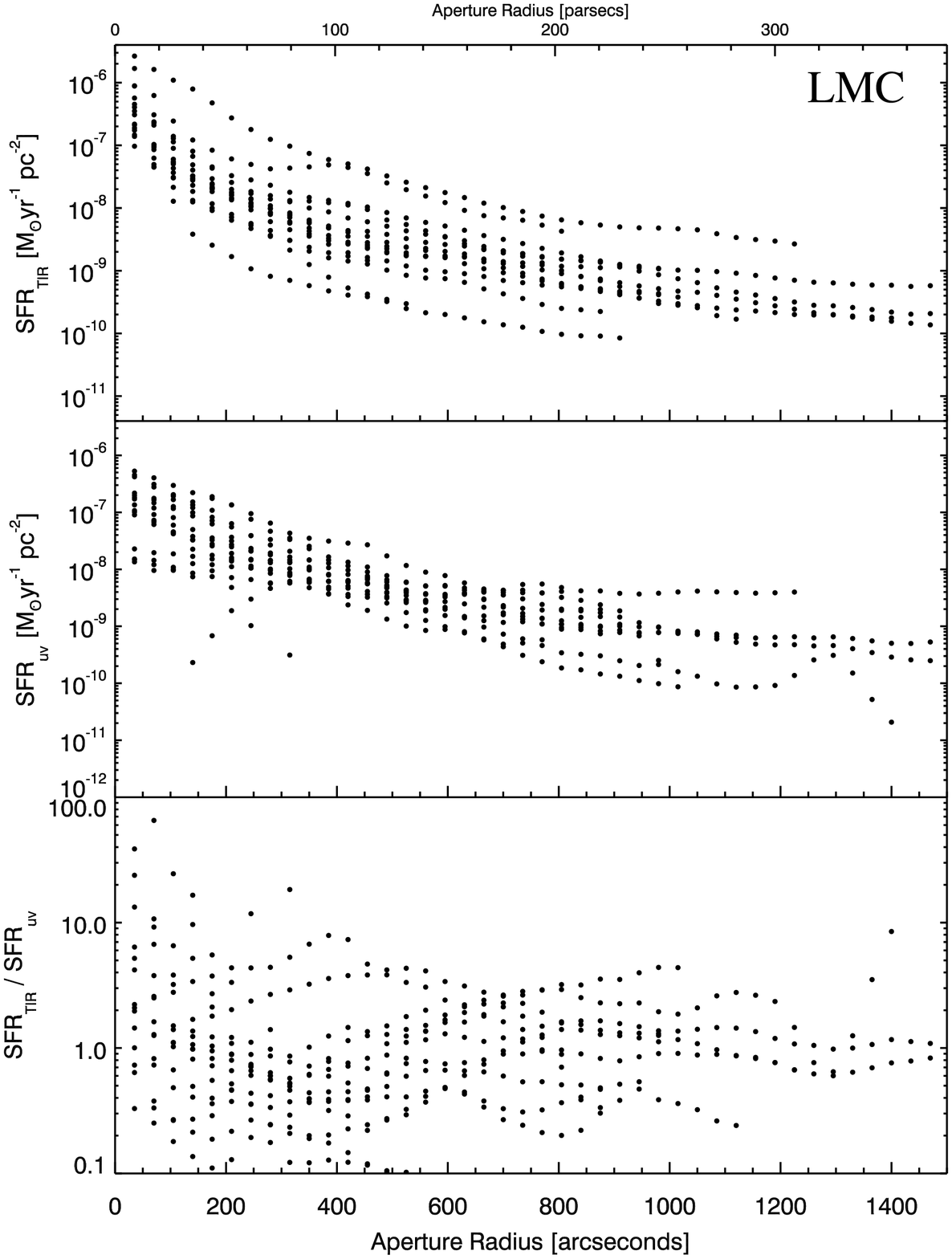}
\caption{Computed radial obscured (TIR--top panel) and unobscured
(UV--middle panel) SFR densities ($M_{\odot}$ yr$^{-1}$ pc$^{-2}$) for
the LMC {\HII} regions, using the SFR equations~1 and 4 from
\citet{kenn98}.  The bottom panel is the ratio of the obscured
TIR-derived SFRs to the unobscured UV-derived SFRs.  The radial units
are in arcseconds with spatial radial units of parsecs on the top
axis.}\label{fig:SFR_uv}
\end{figure}

We do not have UV data of the SMC, but we do have UV data of the LMC.
For the LMC, we can do the same analysis of unobscured SFRs via UV as
we did with {\Ha} (see Figure~\ref{fig:SFR_uv}).  In general, the
UV-derived SFR densities are similar to the TIR-derived SFR densities
in that they are both centrally peaked.  The spatial congruence is
probably a resolution effect because it is contrary to the higher
resolution work of \citep{rela09} who found UV emission offset from IR
emission.  There is no discernable trend in the ratio of obscured to
unobscured SFRs in the bottom panel.

\subsubsection{Obscured SFR Indicators}\label{sec_SFRcons}
\hspace{\parindent}It is well argued that the TIR is the best obscured
SFR indicator \citep{kenn98}, and, indeed, \citet{kenn09} find that
TIR is a better star formation indicator than 8 or 24 $\mu$m.
However, if an observer does not have multiple IR bands then it may be
difficult to make an accurate calculation of TIR.  Thus, a
monochromatic star formation indicator that can reliably recover the
TIR over a large spatial scale is valuable.

We substitute the TIR luminosity in Equation~\ref{EQ:SFR} with the 8,
24, 70, and 160~$\mu$m luminosities by using the average luminosity of
a given band, normalized by TIR.  The resulting SFR equation contains
observable quantities.  These equations take the same form as
Eq.~\ref{EQ:SFR}, but there are band specific constants added ($C_8$,
$C_{24}$, $C_{70}$, and $C_{160}$) that are from the factor $L_{\nu}
L_{TIR}^{-1}$ used to transform $L_{TIR}$ to $L_{\nu}$.  From these
constants we compute the SFRs ($M_{\odot}$ yr$^{-1}$) using the
measured 8, 24, 70, or 160~$\mu$m luminosities (ergs s$^{-1}$),
\begin{equation}
\mbox{SFR}=4.5\times10^{-44}~C_{\nu}~L_{\nu},
\end{equation}
where SFR is in units of solar mass per year, $4.5\times10^{-44}$ is
the constant from \citet{kenn98}, $C_{\nu}$ is the IR band specific
constant, and $L_{\nu}$ is the observed luminosity in ergs per second.

In Table~\ref{tab:SFR_Const}, we show the aperture size-dependent
constants used for the aperture radial sizes from 10 to 300 pc for the
LMC (top rows) and from 10 to 200 pc for the SMC (bottom rows).
Listed in the columns are the size of the aperture radius in parsecs,
the SFR equation constants, and the number of {\HII} regions included
in the calculation.  30~Doradus was not included in these calculations
due to the saturation of 24 and 160~$\mu$m.

\begin{deluxetable}{rrrrrr}
\tabletypesize{\scriptsize}
\tablecolumns{6}
\tablewidth{0pt}
\tablecaption{SFR Equation Average Constants\label{tab:SFR_Const}}
\tablehead{
\colhead{Radius}            &
\colhead{$C_8$}             &
\colhead{$C_{24}$}          &
\colhead{$C_{70}$}          &
\colhead{$C_{160}$}         &
\colhead{$N$}           \\
\colhead{(pc$^{a}$)}        &
\colhead{}                  &
\colhead{}                  &
\colhead{}                  &
\colhead{}                  &
\colhead{}}
\startdata
\multicolumn{6}{c}{LMC} \\ 
\hline
10        & $8.84\pm1.80$          & $5.07\pm0.80$          & $2.19\pm0.15$         & $6.11\pm2.40$         & 15        \\
20        & $8.30\pm1.79$          & $5.89\pm0.91$          & $2.10\pm0.11$         & $5.42\pm1.70$         & 15        \\
50        & $6.65\pm1.22$          & $7.92\pm1.43$          & $2.08\pm0.12$         & $4.05\pm0.93$         & 15        \\
100       & $5.79\pm0.97$          & $9.85\pm1.98$          & $2.11\pm0.15$         & $3.39\pm0.74$         & 15        \\
200       & $5.09\pm0.83$          & $12.46\pm2.46$         & $2.18\pm0.17$         & $2.83\pm0.65$         & 13        \\
300       & $4.57\pm0.27$          & $14.36\pm2.16$         & $2.23\pm0.11$         & $2.49\pm0.32$         & 4         \\
\hline
\multicolumn{6}{c}{SMC} \\ 
\hline
10        & $16.64\pm3.14$         & $8.72\pm3.33$          & $2.22\pm0.18$        & $4.46\pm1.62$          & 16        \\
20        & $16.45\pm3.26$         & $10.16\pm3.45$         & $2.18\pm0.15$        & $4.09\pm1.32$          & 16        \\
50        & $16.70\pm4.40$         & $13.55\pm4.75$         & $2.17\pm0.15$        & $3.53\pm0.87$          & 16        \\
100       & $16.49\pm5.45$         & $16.33\pm5.63$         & $2.19\pm0.15$        & $3.20\pm0.63$          & 16        \\
200       & $18.82\pm3.18$         & $11.93\pm4.61$         & $2.18\pm0.19$        & $3.49\pm1.05$          & 2        
\enddata

\tablecomments{Constants to the SFR equations are calculated for each
aperture radius (parsecs).  The top values are for the LMC and the
bottom values are for the SMC.  The constants are defined in
Section~\ref{sec_SFRcons}.  30~Doradus (N157) has been removed from
this analysis due to saturation in the core in 24 and 160 $\mu$m.  The
uncertainties are given as $\pm1\sigma$.}
\tablenotetext{\mbox{a}}{Using the measured distances of $\sim$52,000
pc to the LMC \citep{szew08} and $\sim$60,500 pc to the SMC
\citep{hild05}.}
\end{deluxetable}

The SFR equations for each of the observable bands are dependent on
the physical scale of the instrumental PSF, particularly for the 8,
24, and 160~$\mu$m luminosities.  The deviation of 8 and 24~$\mu$m
between the LMC and SMC, as discussed in section~\ref{radial}, can be
seen in the differences of $C_{8}$ and $C_{24}$ in
Table~\ref{tab:SFR_Const}.  The 70~$\mu$m band specific constants,
$C_{70}$, do not vary between the LMC and SMC, or with radius, within
the $\pm1\sigma$ standard deviations.  Any SFRs derived from the 8,
24, or 160~$\mu$m luminosities should be used with caution.

The \citet{calz07} SFR formalism is commonly used in the literature
\citep[e.g.,][]{wu08,gala09}.  Using 24~$\mu$m, we compare the SFRs
calculated from our \citet{kenn98} modified method with the
extinction-corrected method of \citet[][see their Equation
(9)]{calz07} for apertures of increasing size.  The SFR equation in
\citet{calz07} is computed by finding a best-fit line through a plot
of {\Pa} luminosity versus 24~$\mu$m luminosity, but using the highest
metallicity galaxies in their sample.  In the inner radii ($\sim10$
pc) of our sample, the results using the formalism of \citet{calz07}
produce SFRs that are greater by a factor of 2 in the LMC and 1.7 in
the SMC.  The SFRs between the two methods slowly equilibrate with
radius until they are nearly identical at a radius of $\sim280$ pc in
the LMC and $\sim140$ pc in the SMC.  The results diverge again in the
outskirts of the {\HII} regions.  At 200 pc, the \citet{calz07} method
produces SFRs that are $\sim2/3$ the value of our SFRs in the LMC and
$\sim3/4$ the value of our SFRs in the SMC.  It is reasonable to
expect our SFRs to be lower near the center of {\HII} regions because
we do not take into account the unextincted UV photons that is in the
\citet{calz07} formalism.  The LMC and SMC differences are likely
because the formalism of \citet{calz07} assumes high metallicity
systems, whereas, the LMC and SMC have different, and low,
metallicities.  Thus, the SFR formalism of \citet{calz07} is not as
accurate for lower metallicity systems, but our derived SFR formalism
is less accurate if there is a large percentage of unextincted UV
photons.  Thus, we stress that our star formation formalism provides
obscured SFRs.

\citet{wu08} also use 24~$\mu$m as a star formation indicator.  They
do a study of 28 compact dwarf galaxies and calculate SFRs using the
24~$\mu$m SFR recipe from \citet{wu05}, which was derived using the
observed correlations between {\Ha}, 1.4~GHz, and 24~$\mu$m
luminosities.  Unlike the \citet{calz07} sample, \citet{wu05} include
both metal-poor and metal-rich galaxies.  The SFRs used via the
\citet{wu05} method better match the SFRs estimated using 1.4~GHz, and
\citet{wu08} note that this method derives SFRs that are up to several
times larger than the SFRs derived using the formalism from
\citet{calz07}.  Comparing our \citet{kenn98} modified 24~$\mu$m
method of calculating radial cumulative SFRs with the results of
\citet{calz07} and \citet{wu05}, we find that the SFRs measured using
our method are $\sim35\%-57\%$ the values found using the \citet{wu05}
method, for all radii.  This differs from the results of
\citet{calz07}, where our SFRs became relatively greater at larger
radii.  Thus, SFRs calculated at larger radii using our formalism lie
between those giving higher SFRs using the \citet{wu05} method and the
lower SFRs calculated using the \citet{calz07} method.  These results
are consistent because the LMC and SMC have lower metallicities than
the normal star forming galaxies used to derive the \citet{calz07}
formalism.

By observing 22 starburst galaxies, \citet{bran06} explore the use of
aromatics as star formation indicators.  They find a correlation
between the aromatic 6.2~$\mu$m emission and the TIR, where the TIR
calculated is uncertain to within a factor of 2.  Using our measured
aromatic 8~$\mu$m emission and the uncertainties in $C_{8}$ in
Table~\ref{tab:SFR_Const}, we can derive TIRs that are uncertain by a
factor of $\sim1.2-1.5$ in the LMC and $\sim1.2-2.0$ in the SMC.  The
differences in our uncertainties are due to the changing dispersions
of $C_{8}$ between the LMC and SMC and between radii.  The SMC
correlations are more uncertain in our sample than the LMC
correlations.  Our results are comparable or better than the results
of \citet{bran06}.  Our measurements do not suffer as much from the
introduction of diffuse dust emission because we center our
observations around star forming areas.  The \citet{bran06} dataset
uses the \textit{Spitzer} IRS slits, and the smallest linear scales
they probe are about 200 pc, whereas, we probe down to $\approx20$ pc.
\citet{bran06} note that any local variations may be ``averaged out''
in their measurements.

The results of \citet{bran06} illustrate some of the difficulties in
estimating TIR associated with star formation when averaging over an
entire galaxy.  Our data are acquired by observing around the central
ionizing sources of the {\HII} regions.  Thus, a caveat is that our
results are more applicable to star forming regions and not entire
galaxies, where the IR emission of cold diffuse dust is more
prominent.  As an example, the average $C_{70}$ value for the LMC and
SMC is 2.17, which corresponds to $L_{TIR} = 0.46 \times L_{70}$.  The
derived values of $C_{70}$ using the total LMC and SMC galactic
apertures are 2.56 and 2.22, respectively.  The SMC value is on the
high end measured in Table~\ref{tab:SFR_Const}, and the total LMC
$C_{70}$ constant is larger, beyond the $\pm1\sigma$ standard
deviations, of those measured for the average {\HII} regions at any
radii in Table~\ref{tab:SFR_Const}.  The uncertainty on the final
calculated SFR after integrating over an entire galaxy will depend
upon how much of the IR photons originate from star forming regions
and how much are due to photons from the general stellar population.

\subsubsection{The 70~$\mu$m Star Formation Proxy}\label{sec:70sfr}
\hspace{\parindent} The results in Table~\ref{tab:SFR_Const} indicate
that 70~$\mu$m is the most suitable obscured SFR indicator.  The
$C_{70}$ constant remains the same value for all radii and between the
LMC and SMC, within the $1\sigma$ standard deviations.  For these
reasons, we calculate one mean value of $C_{70}$ using all of the
$C_{70}$ values at every radius from 10 to 300 pc, including both the
LMC and SMC.  From this, we have computed a monochromatic 70 $\mu$m
SFR equation using the formalism from Equation (4) of \citet{kenn98}
and the mean value of $C_{70}$.  The monochromatic obscured SFR
equation is
\begin{equation}\label{EQ:SFR_FINAL}
\mbox{SFR} = 9.7(0.7)\times10^{-44}~L_{70},
\end{equation}
where SFR is in units of solar masses per year and $L_{70}$ is in
units of ergs per second.  The $1\sigma$ standard deviation in
Equation~(\ref{EQ:SFR_FINAL}) is from the uncertainties in the spread
of $C_{70}$.  The mean of the $C_{70}$ radial averages is $2.17$ and
the 1~$\sigma$ standard deviation (0.15) comes from the mean of the
$C_{70}$ radial standard deviations.  The $1\sigma$ standard deviation
of $C_{70}$ is only $\sim7\%$ of the mean value of $C_{70}$.  As is
the case for the original obscured SFR equation from \citet{kenn98}
(see Eq.~\ref{EQ:SFR}), this equation works as a good approximation
for the total SFR if the area measured is almost entirely obscured by
dust.

We calculate the total obscured SFR of the LMC, using
Equation~\ref{EQ:SFR_FINAL} and the cumulative 70~$\mu$m flux of the
LMC (see section~\ref{res:SEDs}).  The LMC obscured SFR of 0.17
($M_{\odot}$ yr$^{-1}$) agrees well with the results of \citet{harr09}
who estimate the total mean LMC SFR to be 0.2 ($M_{\odot}$ yr$^{-1}$),
with variations over the last five billion years of up to a factor of
two.

The usefulness of Equation~(\ref{EQ:SFR_FINAL}) extends to {\HII}
regions measured with aperture radii anywhere from 10 to 300 pc.  Care
should be taken if measuring sizes outside of these ranges.  On very
small physical scales our relation, as all others, should break down
because the effects of individual protostars and stars will become
important \citep{inde08}.  Furthermore, the effects of hotter {\HII}
regions, like 30~Doradus, may result in a radially dependent 70/TIR
ratio that would introduce errors into our formalism at lower radii
(see the discussion in section~\ref{radial}).  There is also the
possibility that the luminosities of smaller {\HII} regions are
different than expected from a presumed IMF due to small number
statistics \citep{oey98}.  For the same reason, \citet{inde08} argue
that measurements of many smaller clusters will produce SFRs that are
too low since they are not fully sampling the stellar IMF.

Equation~(\ref{EQ:SFR_FINAL}) also may not hold at very large radial
distances because of dust emission excited from older (colder) stellar
populations contaminating the IR SEDs.  Further work must be done to
model the contribution of older stellar populations to the 70~$\mu$m
emission, particularly in the outer regions.  A careful background
subtraction will help mitigate any contamination, and a modest annulus
of radius $\sim100$ pc is all that is required to obtain most of the
bolometric dust emission in the LMC and SMC {\HII} regions (see
section~\ref{size}).  Furthermore, our equation is valid over the wide
range of {\HII} region luminosities and sizes probed in this work.
The low end of the Magellanic Cloud luminosity functions includes
clusters with luminosities on the order of the Orion Nebula
\citep{kenn86}, approximately equivalent to N90 in the SMC, while the
high end covers the massive 30~Doradus {\HII} region.  However, this
research would benefit from a sample of {\HII} regions in a larger
galaxy sample so as to more broadly explore environmental effects.

\section{CONCLUSIONS}
\label{conc}
\hspace{\parindent}We performed UV, {\Ha}, and IR aperture/annulus
photometry of 16 LMC and 16 SMC {\HII} regions at the common
resolution of the MIPS 160~$\mu$m $40\arcsec$ {\fwhm} PSF.  With the
aperture photometry, we computed IR SEDs using the total cumulative
\textit{Spitzer} IRAC (3.6, 4.5, 8~$\mu$m) and MIPS (24, 70,
160~$\mu$m) fluxes of the LMC, SMC, and each {\HII} region.  The
cumulative SED of the LMC is brighter than the SMC, and their shapes
are consistent with IR SEDs of the Magellanic Clouds in the
literature.  The SMC {\HII} regions have cumulative fluxes that are
weaker, by about an order of magnitude, than the LMC {\HII} regions.
We argue that this is expected from what is known about the LMC and
SMC {\HII} region luminosity functions.

We compute the IR bolometric flux (TIR) for each {\HII} region
aperture/annulus by integrating under the linear extrapolation of the
8, 24, 70, and 160~$\mu$m fluxes.  We include the input from the
Rayleigh--Jeans tail of a computed dust MBB by integrating in steps of
0.25~\AA, from 160 to $500~\mu$m.

Using annulus photometry of the {\HII} regions, at distances of
$\sim10$ up to $\sim400$ pc from their cores, we plot the radial
8~$\mu$m, 24~$\mu$m, 70~$\mu$m, 160~$\mu$m and TIR surface brightness
profiles of the 32 {\HII} regions.  The surface brightness profiles
peak near the center of the {\HII} regions for all IR bands and TIR.
We argue that resolution effects may play a role in reducing the SMC
{\HII} region surface brightness profiles relative to the LMC {\HII}
region surface brightness profiles.  The surface brightness profiles
are used to compute the sizes of each {\HII} region, using the TIR,
8~$\mu$m, 24~$\mu$m, 70~$\mu$m, and 160~$\mu$m emission.  The typical
sizes measured have large scatter and are $\approx75$ pc for 8~$\mu$m,
$\approx50$ pc for 24~$\mu$m, $\approx70$ pc for both 70~$\mu$m and
TIR, and $\approx90$ pc for 160~$\mu$m.  The large scatter for each
size measurement likely stems from sampling different parts of the LMC
and SMC luminosity functions.

We compute normalized (by TIR) IR SEDs for each {\HII} region annulus.
The radial SEDs nearly all peak around 70~$\mu$m at all radii for
every {\HII} region, out to $\sim400$ pc.  The result of this is that
it gives four favorable reasons to choose 70~$\mu$m as a monochromatic
obscured star formation indicator, and they are (1) 70~$\mu$m emission
most closely traces the size of {\HII} regions as found using the TIR
($\approx70$ pc in radius); (2) 70~$\mu$m flux, normalized by TIR,
remains nearly constant with radius
($L_{70}\approx45\%~L_{\mbox{TIR}}$) from $\sim10$ to 400 pc; (3)
70~$\mu$m flux, normalized by TIR, has the smallest fractional error
($0.05-0.12$ out to 220 pc); and (4) 70 $\mu$m flux, normalized by
TIR, does not systematically differ between the LMC and SMC.

Radial SFR density plots, using the obscured indicator (TIR) and
unobscured indicators (UV, {\Ha}), are computed using the SFR recipes
of \citet{kenn98}.  The computed SFRs for all indicators peak near the
center of each {\HII} region.  The SFRs for the LMC are larger than
for the SMC due to the larger luminosities of the LMC {\HII} regions.
Within the limits of our resolution, the spatial similarities between
UV, {\Ha}, and 24~$\mu$m are similar to results in the literature.

A modified version of the obscured SFR equation from \citet{kenn98} is
created that depends on a conversion factor from TIR to a
monochromatic IR band.  The conversion factors for the 8, 24, and
160~$\mu$m bands all have large uncertainties and are dependent on
radius and/or host galaxy.  The conversion factor for 70~$\mu$m is
constant, within the low uncertainties, at distances from 10 to 300
pc, and between the LMC and SMC.  We produce a final modified obscured
SFR equation using a single averaged 70~$\mu$m conversion constant.
This \citet{kenn98} modified equation is applicable for
Magellanic-like {\HII} regions and for aperture sizes of $10-300$ pc
radius.

The LMC and SMC have different environmental properties than those
galaxies of earlier Hubble types.  The dust emission observed around
Magellanic Cloud {\HII} regions may not be indicative of dust emission
around {\HII} regions in other galaxies.  The analysis in this work
would be greatly benefited by including {\HII} regions in a larger
sample of galaxies.

By analyzing the 30~Doradus IR properties, we found possible caveats
in using 70~$\mu$m as an obscured star formation indicator.  If {\HII}
regions in a particular galaxy have, on average, similar fluxes as
30~Doradus, then the 70~$\mu$m emission, relative to the TIR, will
exhibit more scatter.  However, we do not expect this to be the case
because 30~Doradus is on the extreme high end of {\HII} region
{\Ha}-derived luminosity functions \citep{kenn89}.

Another general caveat to mention is that this work applies directly
to {\HII} regions and their immediate surroundings.  It is general
practice for those studying SFRs over entire galaxies to use the SFR
recipes calculated from {\HII} regions, despite the added
uncertainties of going from a local star forming region to an entire
galaxy.  The results from this work, applied to entire galaxies, will
be subject to similar uncertainties.

The recent launch of the \textit{HERSCHEL} telescope will greatly
increase our understanding of star formation in galaxies.  At 3.5~m,
not only is \textit{HERSCHEL} the largest space telescope ever
launched, but the PACS photometer, with the ability to observe at
75~$\mu$m, 110~$\mu$m, or 175~$\mu$m, will provide a measure of {\HII}
region SEDs at wavelengths near their peak emissions.  The SPIRE
camera will observe at 250~$\mu$m, 350~$\mu$m, and 500~$\mu$m (albeit
with a lower resolution than PACS) and can provide a better constraint
on the Rayleigh--Jeans side of the dust blackbody emission.  NASA's
2.5~m \textit{SOPHIA} telescope will also have a suite of
high-resolution instruments capable of observing star forming regions
throughout the IR, including near the expected peaks of the {\HII}
region SEDs.

This work is based on observations made with the \textit{Spitzer Space
Telescope}, which is operated by the Jet Propulsion Laboratory,
California Institute of Technology, under NASA contract 1407.  The
SAGE-LMC project has been supported by NASA/Spitzer grant 1275598 and
Meixner's efforts have had additional support from NASA NAG5-12595.

\acknowledgments

\textit{Facilities:} \facility{\textit{Spitzer}(IRAC, MIPS)}

\bibliography{bibliography}

\begin{thebibliography}{83}
\expandafter\ifx\csname natexlab\endcsname\relax\def\natexlab#1{#1}\fi
\expandafter\ifx\csname href\endcsname\relax
  \def\href#1#2{}\fi
\expandafter\ifx\csname urllinklabel\endcsname\relax
  \def\urllinklabel{[LINK]}\fi
\expandafter\ifx\csname adsurllinklabel\endcsname\relax
  \def\adsurllinklabel{[ADS]}\fi

\bibitem[{{Aguirre} {et~al.}(2003){Aguirre}, {Bezaire}, {Cheng}, {Cottingham},
  {Cordone}, {Crawford}, {Fixsen}, {Knox}, {Meyer}, {Norgaard-Nielsen},
  {Silverberg}, {Timbie}, \& {Wilson}}]{agui03}
{Aguirre}, J.~E., {et~al.} 2003, \apj, 596, 273


\bibitem[{{Bernard} {et~al.}(2008){Bernard}, {Reach}, {Paradis}, {Meixner},
  {Paladini}, {Kawamura}, {Onishi}, {Vijh}, {Gordon}, {Indebetouw}, {Hora},
  {Whitney}, {Blum}, {Meade}, {Babler}, {Churchwell}, {Engelbracht}, {For},
  {Misselt}, {Leitherer}, {Cohen}, {Boulanger}, {Frogel}, {Fukui}, {Gallagher},
  {Gorjian}, {Harris}, {Kelly}, {Latter}, {Madden}, {Markwick-Kemper},
  {Mizuno}, {Mizuno}, {Mould}, {Nota}, {Oey}, {Olsen}, {Panagia},
  {Perez-Gonzalez}, {Shibai}, {Sato}, {Smith}, {Staveley-Smith}, {Tielens},
  {Ueta}, {Van Dyk}, {Volk}, {Werner}, \& {Zaritsky}}]{bern08}
{Bernard}, J.-P., {et~al.} 2008, \aj, 136, 919


\bibitem[{{Bernard-Salas} {et~al.}(2009){Bernard-Salas}, {Spoon},
  {Charmandaris}, {Lebouteiller}, {Farrah}, {Devost}, {Brandl}, {Wu}, {Armus},
  {Hao}, {Sloan}, {Weedman}, \& {Houck}}]{sala09}
{Bernard-Salas}, J., {et~al.} 2009, \apjs, 184, 230


\bibitem[{{Bica} \& {Schmitt}(1995)}]{bica95}
{Bica}, E.~L.~D. \& {Schmitt}, H.~R. 1995, \apjs, 101, 41


\bibitem[{{Brandl} {et~al.}(2006){Brandl}, {Bernard-Salas}, {Spoon}, {Devost},
  {Sloan}, {Guilles}, {Wu}, {Houck}, {Weedman}, {Armus}, {Appleton}, {Soifer},
  {Charmandaris}, {Hao}, {Higdon}, {Marshall}, \& {Herter}}]{bran06}
{Brandl}, B.~R., {et~al.} 2006, \apj, 653, 1129


\bibitem[{{Calzetti} {et~al.}(2007){Calzetti}, {Kennicutt}, {Engelbracht},
  {Leitherer}, {Draine}, {Kewley}, {Moustakas}, {Sosey}, {Dale}, {Gordon},
  {Helou}, {Hollenbach}, {Armus}, {Bendo}, {Bot}, {Buckalew}, {Jarrett}, {Li},
  {Meyer}, {Murphy}, {Prescott}, {Regan}, {Rieke}, {Roussel}, {Sheth}, {Smith},
  {Thornley}, \& {Walter}}]{calz07}
{Calzetti}, D., {et~al.} 2007, \apj, 666, 870


\bibitem[{{Calzetti} {et~al.}(2005){Calzetti}, {Kennicutt}, {Bianchi},
  {Thilker}, {Dale}, {Engelbracht}, {Leitherer}, {Meyer}, {Sosey}, {Mutchler},
  {Regan}, {Thornley}, {Armus}, {Bendo}, {Boissier}, {Boselli}, {Draine},
  {Gordon}, {Helou}, {Hollenbach}, {Kewley}, {Madore}, {Martin}, {Murphy},
  {Rieke}, {Rieke}, {Roussel}, {Sheth}, {Smith}, {Walter}, {White}, {Yi},
  {Scoville}, {Polletta}, \& {Lindler}}]{calz05}
---. 2005, \apj, 633, 871


\bibitem[{{Cohen} {et~al.}(2003){Cohen}, {Staveley-Smith}, \& {Green}}]{cohe03}
{Cohen}, M., {Staveley-Smith}, L., \& {Green}, A. 2003, \mnras, 340, 275


\bibitem[{{Cohen} {et~al.}(1988){Cohen}, {Dame}, {Garay}, {Montani}, {Rubio},
  \& {Thaddeus}}]{cohe88}
{Cohen}, R.~S., {et~al.} 1988, \apjl, 331, L95


\bibitem[{{Dale} {et~al.}(2005){Dale}, {Bendo}, {Engelbracht}, {Gordon},
  {Regan}, {Armus}, {Cannon}, {Calzetti}, {Draine}, {Helou}, {Joseph},
  {Kennicutt}, {Li}, {Murphy}, {Roussel}, {Walter}, {Hanson}, {Hollenbach},
  {Jarrett}, {Kewley}, {Lamanna}, {Leitherer}, {Meyer}, {Rieke}, {Rieke},
  {Sheth}, {Smith}, \& {Thornley}}]{dale05}
{Dale}, D.~A., {et~al.} 2005, \apj, 633, 857


\bibitem[{{Dale} \& {Helou}(2002)}]{dale02}
{Dale}, D.~A. \& {Helou}, G. 2002, \apj, 576, 159


\bibitem[{{Dale} {et~al.}(2001){Dale}, {Helou}, {Contursi}, {Silbermann}, \&
  {Kolhatkar}}]{dale01}
{Dale}, D.~A., {et~al.} 2001, \apj, 549, 215


\bibitem[{{Davies} {et~al.}(1976){Davies}, {Elliott}, \& {Meaburn}}]{davi76}
{Davies}, R.~D., {Elliott}, K.~H., \& {Meaburn}, J. 1976, \memras, 81, 89


\bibitem[{{Degioia-Eastwood}(1992)}]{degi92}
{Degioia-Eastwood}, K. 1992, \apj, 397, 542


\bibitem[{{Desert} {et~al.}(1990){Desert}, {Boulanger}, \& {Puget}}]{dese90}
{Desert}, F., {Boulanger}, F., \& {Puget}, J.~L. 1990, \aap, 237, 215


\bibitem[{{D{\'e}sert} {et~al.}(2008){D{\'e}sert}, {Mac{\'{\i}}as-P{\'e}rez},
  {Mayet}, {Giardino}, {Renault}, {Aumont}, {Beno{\^i}t}, {Bernard},
  {Ponthieu}, \& {Tristram}}]{dese08}
{D{\'e}sert}, F.-X., {et~al.} 2008, \aap, 481, 411


\bibitem[{{Draine} {et~al.}(2007){Draine}, {Dale}, {Bendo}, {Gordon}, {Smith},
  {Armus}, {Engelbracht}, {Helou}, {Kennicutt}, {Li}, {Roussel}, {Walter},
  {Calzetti}, {Moustakas}, {Murphy}, {Rieke}, {Bot}, {Hollenbach}, {Sheth}, \&
  {Teplitz}}]{drai07b}
{Draine}, B.~T., {et~al.} 2007, \apj, 663, 866


\bibitem[{{Draine} \& {Lee}(1984)}]{drai84}
{Draine}, B.~T. \& {Lee}, H.~M. 1984, \apj, 285, 89


\bibitem[{{Draine} \& {Li}(2007)}]{drai07}
{Draine}, B.~T. \& {Li}, A. 2007, \apj, 657, 810


\bibitem[{{Dupac} {et~al.}(2003){Dupac}, {Bernard}, {Boudet}, {Giard},
  {Lamarre}, {M{\'e}ny}, {Pajot}, {Ristorcelli}, {Serra}, {Stepnik}, \&
  {Torre}}]{dupa03}
{Dupac}, X., {et~al.} 2003, \aap, 404, L11


\bibitem[{{Elbaz} \& {Cesarsky}(2003)}]{elba03}
{Elbaz}, D. \& {Cesarsky}, C.~J. 2003, Science, 300, 270


\bibitem[{{Engelbracht} {et~al.}(2007){Engelbracht}, {Blaylock}, {Su}, {Rho},
  {Rieke}, {Muzerolle}, {Padgett}, {Hines}, {Gordon}, {Fadda},
  {Noriega-Crespo}, {Kelly}, {Latter}, {Hinz}, {Misselt}, {Morrison},
  {Stansberry}, {Shupe}, {Stolovy}, {Wheaton}, {Young}, {Neugebauer},
  {Wachter}, {P{\'e}rez-Gonz{\'a}lez}, {Frayer}, \& {Marleau}}]{enge07}
{Engelbracht}, C.~W., {et~al.} 2007, \pasp, 119, 994


\bibitem[{{Engelbracht} {et~al.}(2008){Engelbracht}, {Rieke}, {Gordon},
  {Smith}, {Werner}, {Moustakas}, {Willmer}, \& {Vanzi}}]{enge08}
---. 2008, \apj, 685, 678


\bibitem[{{Fazio} {et~al.}(2004){Fazio}, {Hora}, {Allen}, {Ashby}, {Barmby},
  {Deutsch}, {Huang}, {Kleiner}, {Marengo}, {Megeath}, {Melnick}, {Pahre},
  {Patten}, {Polizotti}, {Smith}, {Taylor}, {Wang}, {Willner}, {Hoffmann},
  {Pipher}, {Forrest}, {McMurty}, {McCreight}, {McKelvey}, {McMurray}, {Koch},
  {Moseley}, {Arendt}, {Mentzell}, {Marx}, {Losch}, {Mayman}, {Eichhorn},
  {Krebs}, {Jhabvala}, {Gezari}, {Fixsen}, {Flores}, {Shakoorzadeh}, {Jungo},
  {Hakun}, {Workman}, {Karpati}, {Kichak}, {Whitley}, {Mann}, {Tollestrup},
  {Eisenhardt}, {Stern}, {Gorjian}, {Bhattacharya}, {Carey}, {Nelson},
  {Glaccum}, {Lacy}, {Lowrance}, {Laine}, {Reach}, {Stauffer}, {Surace},
  {Wilson}, {Wright}, {Hoffman}, {Domingo}, \& {Cohen}}]{fazi04}
{Fazio}, G.~G., {et~al.} 2004, \apjs, 154, 10


\bibitem[{{Fukui} {et~al.}(2008){Fukui}, {Kawamura}, {Minamidani}, {Mizuno},
  {Kanai}, {Mizuno}, {Onishi}, {Yonekura}, {Mizuno}, {Ogawa}, \&
  {Rubio}}]{fuku08}
{Fukui}, Y., {et~al.} 2008, \apjs, 178, 56


\bibitem[{{Galametz} {et~al.}(2009){Galametz}, {Madden}, {Galliano}, {Hony},
  {Schuller}, {Beelen}, {Bendo}, {Sauvage}, {Lundgren}, \& {Billot}}]{gala09}
{Galametz}, M., {et~al.} 2009, \aap, 508, 645


\bibitem[{{Galliano} {et~al.}(2005){Galliano}, {Madden}, {Jones}, {Wilson}, \&
  {Bernard}}]{gall05}
{Galliano}, F., {et~al.} 2005, \aap, 434, 867


\bibitem[{{Gaustad} {et~al.}(2001){Gaustad}, {McCullough}, {Rosing}, \& {Van
  Buren}}]{gaus01}
{Gaustad}, J.~E., {et~al.} 2001, \pasp, 113, 1326


\bibitem[{{Genzel}(1991)}]{genz91}
{Genzel}, R. 1991, {The Physics of Star Formation and Early Stellar Evolution}
  (Dordrecht:Kluwer)


\bibitem[{{Gezari} {et~al.}(1973){Gezari}, {Joyce}, \& {Simon}}]{geza73}
{Gezari}, D.~Y., {Joyce}, R.~R., \& {Simon}, M. 1973, \apjl, 179, L67+


\bibitem[{{Gieles} {et~al.}(2006){Gieles}, {Larsen}, {Bastian}, \&
  {Stein}}]{giel06}
{Gieles}, M., {et~al.} 2006, \aap, 450, 129


\bibitem[{{Gordon} {et~al.}(2000){Gordon}, {Clayton}, {Witt}, \&
  {Misselt}}]{gord00}
{Gordon}, K.~D., {et~al.} 2000, \apj, 533, 236


\bibitem[{{Gordon} {et~al.}(2007){Gordon}, {Engelbracht}, {Fadda},
  {Stansberry}, {Wachter}, {Frayer}, {Rieke}, {Noriega-Crespo}, {Latter},
  {Young}, {Neugebauer}, {Balog}, {Beeman}, {Dole}, {Egami}, {Haller}, {Hines},
  {Kelly}, {Marleau}, {Misselt}, {Morrison}, {P{\'e}rez-Gonz{\'a}lez}, {Rho},
  \& {Wheaton}}]{gord07}
---. 2007, \pasp, 119, 1019


\bibitem[{{Gordon} {et~al.}(2008){Gordon}, {Engelbracht}, {Rieke}, {Misselt},
  {Smith}, \& {Kennicutt}}]{gord08}
---. 2008, \apj, 682, 336


\bibitem[{{Gordon} {et~al.}(2005){Gordon}, {Rieke}, {Engelbracht}, {Muzerolle},
  {Stansberry}, {Misselt}, {Morrison}, {Cadien}, {Young}, {Dole}, {Kelly},
  {Alonso-Herrero}, {Egami}, {Su}, {Papovich}, {Smith}, {Hines}, {Rieke},
  {Blaylock}, {P{\'e}rez-Gonz{\'a}lez}, {Le Floc'h}, {Hinz}, {Latter},
  {Hesselroth}, {Frayer}, {Noriega-Crespo}, {Masci}, {Padgett}, {Smylie}, \&
  {Haegel}}]{gord05}
---. 2005, \pasp, 117, 503


\bibitem[{{Harris} \& {Zaritsky}(2009)}]{harr09}
{Harris}, J. \& {Zaritsky}, D. 2009, \aj, 138, 1243


\bibitem[{{Helou} {et~al.}(2004){Helou}, {Roussel}, {Appleton}, {Frayer},
  {Stolovy}, {Storrie-Lombardi}, {Hurt}, {Lowrance}, {Makovoz}, {Masci},
  {Surace}, {Gordon}, {Alonso-Herrero}, {Engelbracht}, {Misselt}, {Rieke},
  {Rieke}, {Willner}, {Pahre}, {Ashby}, {Fazio}, \& {Smith}}]{helo04}
{Helou}, G., {et~al.} 2004, \apjs, 154, 253


\bibitem[{{Henize}(1956)}]{heni56}
{Henize}, K.~G. 1956, \apjs, 2, 315


\bibitem[{{Hilditch} {et~al.}(2005){Hilditch}, {Howarth}, \&
  {Harries}}]{hild05}
{Hilditch}, R.~W., {Howarth}, I.~D., \& {Harries}, T.~J. 2005, \mnras, 357, 304


\bibitem[{{Hodge}(1974)}]{hodg74}
{Hodge}, P.~W. 1974, \pasp, 86, 845


\bibitem[{{Hodge} \& {Kennicutt}(1983)}]{hodg83}
{Hodge}, P.~W. \& {Kennicutt}, Jr., R.~C. 1983, \aj, 88, 296


\bibitem[{{Indebetouw} {et~al.}(2008){Indebetouw}, {Whitney}, {Kawamura},
  {Onishi}, {Meixner}, {Meade}, {Babler}, {Hora}, {Gordon}, {Engelbracht},
  {Block}, \& {Misselt}}]{inde08}
{Indebetouw}, R., {et~al.} 2008, \aj, 136, 1442


\bibitem[{{Kaufman} {et~al.}(1999){Kaufman}, {Wolfire}, {Hollenbach}, \&
  {Luhman}}]{kauf99}
{Kaufman}, M.~J., {et~al.} 1999, \apj, 527, 795


\bibitem[{{Kennicutt} {et~al.}(2009){Kennicutt}, {Hao}, {Calzetti},
  {Moustakas}, {Dale}, {Bendo}, {Engelbracht}, {Johnson}, \& {Lee}}]{kenn09}
{Kennicutt}, R.~C., {et~al.} 2009, \apj, 703, 1672


\bibitem[{{Kennicutt} \& {Hodge}(1980)}]{kenn80}
{Kennicutt}, R.~C. \& {Hodge}, P.~W. 1980, \apj, 241, 573


\bibitem[{{Kennicutt}(1998)}]{kenn98}
{Kennicutt}, Jr., R.~C. 1998, \araa, 36, 189


\bibitem[{{Kennicutt} {et~al.}(2007){Kennicutt}, {Calzetti}, {Walter}, {Helou},
  {Hollenbach}, {Armus}, {Bendo}, {Dale}, {Draine}, {Engelbracht}, {Gordon},
  {Prescott}, {Regan}, {Thornley}, {Bot}, {Brinks}, {de Blok}, {de Mello},
  {Meyer}, {Moustakas}, {Murphy}, {Sheth}, \& {Smith}}]{kenn07}
{Kennicutt}, Jr., R.~C., {et~al.} 2007, \apj, 671, 333


\bibitem[{{Kennicutt} {et~al.}(1989){Kennicutt}, {Edgar}, \& {Hodge}}]{kenn89}
{Kennicutt}, Jr., R.~C., {Edgar}, B.~K., \& {Hodge}, P.~W. 1989, \apj, 337, 761


\bibitem[{{Kennicutt} \& {Hodge}(1986)}]{kenn86}
{Kennicutt}, Jr., R.~C. \& {Hodge}, P.~W. 1986, \apj, 306, 130


\bibitem[{{Kim} {et~al.}(1999){Kim}, {Dopita}, {Staveley-Smith}, \&
  {Bessell}}]{kim99}
{Kim}, S., {et~al.} 1999, \aj, 118, 2797


\bibitem[{{Lebouteiller} {et~al.}(2007){Lebouteiller}, {Brandl},
  {Bernard-Salas}, {Devost}, \& {Houck}}]{lebo07}
{Lebouteiller}, V., {et~al.} 2007, \apj, 665, 390


\bibitem[{{Li} \& {Draine}(2001)}]{li01}
{Li}, A. \& {Draine}, B.~T. 2001, \apj, 554, 778


\bibitem[{{Livanou} {et~al.}(2007){Livanou}, {Gonidakis}, {Kontizas}, {Klein},
  {Kontizas}, {Kester}, {Fukui}, {Mizuno}, \& {Tsalmantza}}]{liva07}
{Livanou}, E., {et~al.} 2007, \aj, 133, 2179


\bibitem[{{Martin} {et~al.}(2005){Martin}, {Fanson}, {Schiminovich},
  {Morrissey}, {Friedman}, {Barlow}, {Conrow}, {Grange}, {Jelinsky},
  {Milliard}, {Siegmund}, {Bianchi}, {Byun}, {Donas}, {Forster}, {Heckman},
  {Lee}, {Madore}, {Malina}, {Neff}, {Rich}, {Small}, {Surber}, {Szalay},
  {Welsh}, \& {Wyder}}]{mart05}
{Martin}, D.~C., {et~al.} 2005, \apjl, 619, L1


\bibitem[{{Massey} \& {Hunter}(1998)}]{mass98}
{Massey}, P. \& {Hunter}, D.~A. 1998, \apj, 493, 180


\bibitem[{{Meixner} {et~al.}(2006){Meixner}, {Gordon}, {Indebetouw}, {Hora},
  {Whitney}, {Blum}, {Reach}, {Bernard}, {Meade}, {Babler}, {Engelbracht},
  {For}, {Misselt}, {Vijh}, {Leitherer}, {Cohen}, {Churchwell}, {Boulanger},
  {Frogel}, {Fukui}, {Gallagher}, {Gorjian}, {Harris}, {Kelly}, {Kawamura},
  {Kim}, {Latter}, {Madden}, {Markwick-Kemper}, {Mizuno}, {Mizuno}, {Mould},
  {Nota}, {Oey}, {Olsen}, {Onishi}, {Paladini}, {Panagia}, {Perez-Gonzalez},
  {Shibai}, {Sato}, {Smith}, {Staveley-Smith}, {Tielens}, {Ueta}, {van Dyk},
  {Volk}, {Werner}, \& {Zaritsky}}]{meix06}
{Meixner}, M., {et~al.} 2006, \aj, 132, 2268


\bibitem[{{Mennella} {et~al.}(1998){Mennella}, {Brucato}, {Colangeli},
  {Palumbo}, {Rotundi}, \& {Bussoletti}}]{menn98}
{Mennella}, V., {et~al.} 1998, \apj, 496, 1058


\bibitem[{{Oey} \& {Clarke}(1998)}]{oey98}
{Oey}, M.~S. \& {Clarke}, C.~J. 1998, \aj, 115, 1543


\bibitem[{{Oey} {et~al.}(2003){Oey}, {Parker}, {Mikles}, \& {Zhang}}]{oey03}
{Oey}, M.~S., {et~al.} 2003, \aj, 126, 2317


\bibitem[{{Peeters} {et~al.}(2002){Peeters}, {Mart{\'{\i}}n-Hern{\'a}ndez},
  {Damour}, {Cox}, {Roelfsema}, {Baluteau}, {Tielens}, {Churchwell}, {Kessler},
  {Mathis}, {Morisset}, \& {Schaerer}}]{peet02}
{Peeters}, E., {et~al.} 2002, \aap, 381, 571


\bibitem[{{Puget} \& {Leger}(1989)}]{puge89}
{Puget}, J.~L. \& {Leger}, A. 1989, \araa, 27, 161


\bibitem[{{Reach} {et~al.}(2005){Reach}, {Megeath}, {Cohen}, {Hora}, {Carey},
  {Surace}, {Willner}, {Barmby}, {Wilson}, {Glaccum}, {Lowrance}, {Marengo}, \&
  {Fazio}}]{reac05}
{Reach}, W.~T., {et~al.} 2005, \pasp, 117, 978


\bibitem[{{Rela{\~n}o} \& {Kennicutt}(2009)}]{rela09}
{Rela{\~n}o}, M. \& {Kennicutt}, R.~C. 2009, \apj, 699, 1125


\bibitem[{{Rieke} {et~al.}(2004){Rieke}, {Young}, {Engelbracht}, {Kelly},
  {Low}, {Haller}, {Beeman}, {Gordon}, {Stansberry}, {Misselt}, {Cadien},
  {Morrison}, {Rivlis}, {Latter}, {Noriega-Crespo}, {Padgett}, {Stapelfeldt},
  {Hines}, {Egami}, {Muzerolle}, {Alonso-Herrero}, {Blaylock}, {Dole}, {Hinz},
  {Le Floc'h}, {Papovich}, {P{\'e}rez-Gonz{\'a}lez}, {Smith}, {Su}, {Bennett},
  {Frayer}, {Henderson}, {Lu}, {Masci}, {Pesenson}, {Rebull}, {Rho}, {Keene},
  {Stolovy}, {Wachter}, {Wheaton}, {Werner}, \& {Richards}}]{riek04}
{Rieke}, G.~H., {et~al.} 2004, \apjs, 154, 25


\bibitem[{{Rubio} {et~al.}(1998){Rubio}, {Barb{\'a}}, {Walborn}, {Probst},
  {Garc{\'{\i}}a}, \& {Roth}}]{rubi98}
{Rubio}, M., {et~al.} 1998, \aj, 116, 1708


\bibitem[{{Salpeter}(1955)}]{salp55}
{Salpeter}, E.~E. 1955, \apj, 121, 161


\bibitem[{{Shetty} {et~al.}(2009){Shetty}, {Kauffmann}, {Schnee}, \&
  {Goodman}}]{shet09}
{Shetty}, R., {et~al.} 2009, \apj, 696, 676


\bibitem[{{Smith} {et~al.}(1987){Smith}, {Cornett}, \& {Hill}}]{smit87}
{Smith}, A.~M., {Cornett}, R.~H., \& {Hill}, R.~S. 1987, \apj, 320, 609


\bibitem[{{Smith} \& {Brooks}(2007)}]{smit07}
{Smith}, N. \& {Brooks}, K.~J. 2007, \mnras, 379, 1279


\bibitem[{{Snider} {et~al.}(2009){Snider}, {Hester}, {Desch}, {Healy}, \&
  {Bally}}]{snid09}
{Snider}, K.~D., {et~al.} 2009, \apj, 700, 506


\bibitem[{{Stansberry} {et~al.}(2007){Stansberry}, {Gordon}, {Bhattacharya},
  {Engelbracht}, {Rieke}, {Marleau}, {Fadda}, {Frayer}, {Noriega-Crespo},
  {Wachter}, {Young}, {M{\"u}ller}, {Kelly}, {Blaylock}, {Henderson},
  {Neugebauer}, {Beeman}, \& {Haller}}]{stan07}
{Stansberry}, J.~A., {et~al.} 2007, \pasp, 119, 1038


\bibitem[{{Szewczyk} {et~al.}(2008){Szewczyk}, {Pietrzy{\'n}ski}, {Gieren},
  {Storm}, {Walker}, {Rizzi}, {Kinemuchi}, {Bresolin}, {Kudritzki}, \&
  {Dall'Ora}}]{szew08}
{Szewczyk}, O., {et~al.} 2008, \aj, 136, 272


\bibitem[{{Thilker} {et~al.}(2007){Thilker}, {Boissier}, {Bianchi}, {Calzetti},
  {Boselli}, {Dale}, {Seibert}, {Braun}, {Burgarella}, {Gil de Paz}, {Helou},
  {Walter}, {Kennicutt}, {Madore}, {Martin}, {Barlow}, {Forster}, {Friedman},
  {Morrissey}, {Neff}, {Schiminovich}, {Small}, {Wyder}, {Donas}, {Heckman},
  {Lee}, {Milliard}, {Rich}, {Szalay}, {Welsh}, \& {Yi}}]{thil07}
{Thilker}, D.~A., {et~al.} 2007, \apjs, 173, 572


\bibitem[{{Tielens}(2005)}]{tiel05}
{Tielens}, A.~G.~G.~M. 2005, {The Physics and Chemistry of the Interstellar
  Medium} (Cambridge: Cambridge University Press)


\bibitem[{{van den Bergh}(1981)}]{van81}
{van den Bergh}, S. 1981, \aj, 86, 1464


\bibitem[{{van Loon} {et~al.}(2010{\natexlab{a}}){van Loon}, {Oliveira},
  {Gordon}, {Meixner}, {Shiao}, {Boyer}, {Kemper}, {Woods}, {Tielens},
  {Marengo}, {Indebetouw}, {Sloan}, \& {Chen}}]{jacc10}
{van Loon}, J.~T., {et~al.} 2010{\natexlab{a}}, \aj, 139, 68


\bibitem[{{van Loon} {et~al.}(2010{\natexlab{b}}){van Loon}, {Oliveira},
  {Gordon}, {Sloan}, \& {Engelbracht}}]{jacc10b}
---. 2010{\natexlab{b}}, \aj, 139, 1553


\bibitem[{{Walborn} \& {Blades}(1997)}]{walb97}
{Walborn}, N.~R. \& {Blades}, J.~C. 1997, \apjs, 112, 457


\bibitem[{{Walborn} {et~al.}(2002){Walborn}, {Ma{\'{\i}}z-Apell{\'a}niz}, \&
  {Barb{\'a}}}]{walb02}
{Walborn}, N.~R., {Ma{\'{\i}}z-Apell{\'a}niz}, J., \& {Barb{\'a}}, R.~H. 2002,
  \aj, 124, 1601


\bibitem[{{Walter} {et~al.}(2007){Walter}, {Cannon}, {Roussel}, {Bendo},
  {Calzetti}, {Dale}, {Draine}, {Helou}, {Kennicutt}, {Moustakas}, {Rieke},
  {Armus}, {Engelbracht}, {Gordon}, {Hollenbach}, {Lee}, {Li}, {Meyer},
  {Murphy}, {Regan}, {Smith}, {Brinks}, {de Blok}, {Bigiel}, \&
  {Thornley}}]{walt07}
{Walter}, F., {et~al.} 2007, \apj, 661, 102


\bibitem[{{Watson} {et~al.}(2008){Watson}, {Povich}, {Churchwell}, {Babler},
  {Chunev}, {Hoare}, {Indebetouw}, {Meade}, {Robitaille}, \&
  {Whitney}}]{wats08}
{Watson}, C., {et~al.} 2008, \apj, 681, 1341


\bibitem[{{Wu} {et~al.}(2005){Wu}, {Cao}, {Hao}, {Liu}, {Wang}, {Xia}, {Deng},
  \& {Young}}]{wu05}
{Wu}, H., {et~al.} 2005, \apjl, 632, L79


\bibitem[{{Wu} {et~al.}(2008){Wu}, {Charmandaris}, {Houck}, {Bernard-Salas},
  {Lebouteiller}, {Brandl}, \& {Farrah}}]{wu08}
{Wu}, Y., {et~al.} 2008, \apj, 676, 970


\end{thebibliography}

\end{document}